\documentclass[preprint,superscriptaddress]{revtex4-1}
\usepackage{graphicx}
\usepackage{array}
\usepackage{amsfonts}
\usepackage{amssymb,amsmath,multirow,rotate}
\usepackage{mathrsfs}
\usepackage{booktabs}
\usepackage{threeparttable}
\usepackage{multirow}
\usepackage{epsfig}
\usepackage{threeparttable}
\usepackage{chngpage}
\usepackage{latexsym}
\usepackage{dcolumn}
\usepackage{graphics,graphicx,fleqn,epic,eepic}
\usepackage{times}
\usepackage{verbatim}
\usepackage{subfigure}
\usepackage{color}
\usepackage{float}
\usepackage{bm}

\definecolor{red}{rgb}{1,0,0}

\definecolor{blue}{rgb}{0,0,1}

\graphicspath{{./Figs/}}

\begin{document}

\title{Dynamics of ferrofluidic flow in the Taylor-Couette system with a
       small aspect ratio}

\author{Sebastian Altmeyer} 
\affiliation{Institute of Science and Technology Austria 
(IST Austria), 3400 Klosterneuburg, Austria}

\author{Younghae Do}
\affiliation{Department of Mathematics,
KNU-Center for Nonlinear Dynamics, 
Kyungpook National University, Daegu, 702-701, South Korea}

\author{Ying-Cheng Lai}
\affiliation{School of Electrical, Computer and Energy Engineering, 
Arizona State University, Tempe, Arizona, 85287, USA}

\begin{abstract}

{\bf
We investigate fundamental nonlinear dynamics of ferrofluidic Taylor-Couette flow
- flow confined between two concentric independently rotating cylinders - consider
small aspect ratio by solving the ferrohydrodynamical equations, carrying out
systematic bifurcation analysis. Without magnetic field, we find steady flow patterns,
previously observed with a simple fluid, such as those containing {\em normal} one- or two
vortex cells, as well as {\em anomalous} one-cell and twin-cell flow states. However, when
a {\em symmetry-breaking} transverse magnetic field is present, all flow states exhibit
stimulated, finite two-fold mode. Various bifurcations between steady and unsteady
states can occur, corresponding to the transitions between the two-cell and one-cell
states. While unsteady, {\em axially oscillating} flow states can arise, we also detect the
emergence of new unsteady flow states. In particular, we uncover two new states: one
contains only the {\em azimuthally oscillating} solution in the configuration of the twin-cell
flow state, and another a {\em rotating} flow state. Topologically, these flow states are a
limit cycle and a quasiperiodic solution on a two-torus, respectively. Emergence of
new flow states in addition to observed ones with classical fluid, indicates that
richer but potentially more controllable dynamics in ferrofluidic flows, as such
flow states depend on the external magnetic field.}\\

\end{abstract}
\date{\today}
\maketitle

The flow between two concentric differentially rotating cylinders, the 
Taylor-Couette system (TCS), has played a central role in understanding
the various hydrodynamic stabilities~\cite{Tay:1923,ChIo:1994} TCS
has been a paradigm to investigate many fundamental nonlinear dynamical
phenomena in fluid flows. The simplicity 
of the geometry of the system allows for well-controlled experimental 
studies. The vast literature in this area has been built on the TCS 
with a simple fluid. (For convenience, in this paper we call TCS with
a simple fluid the {\em \textquoteleft classical TCS'}.)   
Recently there has been an increasing amount of interest in the
flow dynamics of the TCS with a complex fluid~\cite{AHLL:2010,ReOd:2010,
ReOd:2011,Alt2011,ALD:2013,ADL:2015a,ADL:2015b}. A representative
type of complex fluids is ferrofluids~\cite{Ros:1985}, which are manufactured 
fluids consisting of dispersion of magnetized nanoparticles in a liquid 
carrier. A ferrofluid can be stabilized against agglomeration through 
the addition of a surfactant monolayer onto the particles. In the absence 
of any magnetic field, the nanoparticles are randomly orientated
so that the fluid has zero net magnetization. In this case, 
the nanoparticles alter little the viscosity and the density of the 
fluid. Thus, in the absence of any external field a ferrofluid behaves as
a simple (classical) fluid. However, when a magnetic field of sufficient 
strength is applied, the hydrodynamical properties of the fluid, such as 
the viscosity, can be changed dramatically~\cite{Mc69,Sh72} and the dynamics 
can be drastically altered. Studies indicated that, under a symmetry-breaking 
transverse magnetic field, all flow states in the TCS become intrinsically
{\em three-dimensional}~\cite{AHLL:2010,ReOd:2011,ALD:2013}. As such, a magnetic
field can have a significant influence on the hydrodynamical stability and
the underlying symmetries of the flow states through, e.g., certain 
induced azimuthal modes~\cite{ALD:2013}. Aside this a change in the magnetic field
strength can also induce turbulence~\cite{ADL:2015a}. Ferrofluidic flows have 
wide applications, ranging from gaining insights into the fundamentals of 
geophysical flows through laboratory experiments~\cite{Har06,HaKi:2006} to
the development of microfluidic devices and computer hard drives. For 
example, a recent study demonstrated that ferrofluidic flows in the TCS can 
reverse their directions of rotation spontaneously~\cite{ADL:2015b}, which 
has implications to the phenomenon of geomagnetic reversal~\cite{GR:1995a,
GR:1995b,GR:1996,GCHR:1999,GR:2000,Har:2002,HaKi:2006,Berhanuetal:2007}.

Our study of the ferrofluidic flow states in the TCS with a small aspect 
ratio was motivated by the following considerations. Previous numerical and 
experimental works demonstrated that the effects of end walls are {\em not} 
negligible~\cite{Ben78a,Ben78b,CKM92,AHHAPLP2010} even in the large aspect 
ratio TCS. The walls can thus have a {\em significant} effect on the flow 
dynamics. For the classical TCS or for TCS with a ferrofluid but
without any magnetic field, for small aspect ratio (e.g., $\Gamma \approx 1$) 
the flow dynamics is dominated by the competition between normal and 
anomalous flow states, leading to rich dynamical behaviors~\cite{BeMu:1981,
LMWP:1984,PSCM:1988,PSL:1991,PBE:1992}. Here the term ``{\em normal}'' 
(``{\em anomalous}'') is referred to as a flow state with vortex cells that give 
an inward (outward) flow near each lid in the radial direction. For 
systems of a small height different flows patterns with one ({\em one-cell 
flow state}) or two ({\em two-cell flow state}) Taylor vortex cells were 
detected~\cite{Cli:1983,SPT:2003}. A plausible mechanism for the 
emergence of the flow states is that the vortex cells, independent of 
the normal or anomalous nature of the flow, divide the flow in the axial 
direction. In addition, flows with two identical cells, the so called 
``twin-cell'' flows, were observed~\cite{NaTo:1996}, in which the two 
vortex cells divide the flow in the radial instead of the common axial 
direction. That is, both cells touch the top and the bottom lids. While most
flow states are steady, an unsteady and axially oscillatory flow state was
also experimentally detected~\cite{BSP:1992} and numerically 
demonstrated~\cite{NaTo:1996}. An alternative type of unsteady 
flow states with more complex dynamics~\cite{FWTN:2002} was observed in 
the TCS with an extraordinarily small aspect ratio (e.g., $\Gamma=0.5$),
where the flow can change among two, three, and four cells in a radially 
separating configuration over one period. To summarize briefly, existing
works on the classical TCS demonstrated that complex oscillatory
flow patterns can arise when the aspect ratio of the system is reduced.
An open issue is what types of dynamical behaviors can arise in the flow
patterns in the ferrofluidic TCS, subject to a magnetic field.
 
In this paper, we report the results from a systematic computational study
of the ferrofluidic flow dynamics in the TCS with a small aspect ratio, i.e., 
on the order of unity, which we choose as a bifurcation parameter (The 
radius ratio of the cylinders [inner cylinder radius / outer cylinder radius]
is fixed to $0.5$). Another bifurcation parameter is the Reynolds number
($Re_2=\omega_2 r_2 d/\nu$ see {\bf Methods}) of the outer cylinder.  
Specifically, we set the rotation speed of the inner cylinder so as to 
fix its Reynolds number at $Re_1=250$, and vary the rotation speed of the
outer cylinder. Both end walls confining the TCS are stationary. To 
distinguish from the dynamics of a simple fluid, we apply a symmetry 
breaking, transverse magnetic field. The
main results can be stated as follows. We find that all flow states
exhibit a general feature: they contain a stimulated two-cell mode
\cite{AHHAPLP2010,ReOd:2011,ALD:2012}. As 
the aspect ratio is changed, various bifurcations between steady and 
unsteady flow states can occur, corresponding to the transitions 
between the two-cell and one-cell states. While unsteady, axially oscillating 
flow states similar to those in a simple fluid can occur,
novel unsteady flow states that are not found in the classical TCS
can arise. In particular, we uncover two new states: one that contains only 
the {\em azimuthally} oscillating solution in the configuration of the 
twin-cell flow state, and another a rotating flow state, which correspond
topologically to limit cycle and quasiperiodic solution on a two-torus, 
respectively. Due to the sequence of bifurcations following a symmetry 
breaking bifurcation, the one-cell and twin-cell flow states are 
symmetrically related. We also uncover various regions of bistability 
with the coexistence of one- and two-cell flow states. The emergence of the 
novel flow states in addition to those occurring typically in the classical
TCS suggests that the ferrofluidic TCS can exhibit richer dynamics that
are potentially more controllable due to their dependence on an additional
experimentally adjustable parameter: the magnetic field strength.   
%
\\
\\ \noindent  

{\large\bf Results} 

\paragraph*{Nomenclature.} 
We focus on the flow states in the small aspect-ratio TCS. A common feature
shared by all flow states is that the axisymmetric Fourier mode associated
with the azimuthal wavenumber $m=0$ (see {\bf Methods}) is dominant so that 
the flow states correspond to {\em toroidally closed} solutions. Note that 
ferrofluidic flows dominated by an azimuthally modulated $m=0$ mode differ 
from the classical wavy vortex flow solutions in the absence of any 
magnetic field~\cite{GSS88,GL88,WeLu:1998,CHSAML2009,MSL:2014}, which 
are time-periodic, {\em rotating} 
states that do {\em not} propagate axially. In the presence of a 
transverse magnetic field, all the flow states are fundamentally three 
dimensional with a stimulated $m=2$ mode, leading to steady 
({\em non-rotating}) wavy vortex flows~\cite{AHLL:2010,ReOd:2011,ALD:2013}.
Rotating flows with a finite $m=1$ mode can also arise, so do unsteady 
(oscillatory) flow solutions. A key indicator differentiating various 
flow states is the number of vortex cells present in the annulus, i.e., 
in the $(r,z)$ plane. To take into account all the differences, we use the 
notation $\# \text{\bf Con}_{\text{\bf m}}^{\text{\bf spec}}$ defined in 
Tab.~\ref{tab:names} to distinguish the different flow patterns.
For example, the notion $2N_2^\text{z-osci}$ stands for an unsteady, 
axially oscillating [z-osci] two-cell [2] flow state in the normal configuration [N] 
with a stimulated $m=2$ mode $[_2]$.  
It is worth mentioning that all calculated wavy flows are {\em stable}.
However, for the parameter regimes considered the Taylor-vortex flow (TVF) 
solutions are unstable. The magnetic field strength can be characterized by 
the Niklas parameter (see {\bf Methods}). In the present work we consider a 
transverse field with fixed parameter $s_x=0.6$. The velocity and vorticity 
fields are ${\bm u}=(u,v,w)$ and $\nabla \times \bm{u} = (\xi,\eta,\zeta)$, 
respectively.

\begin{table}{}
\begin{center}
\begin{tabular}{|c|c|c|} \hline
\multicolumn{3}{|c|}{$\# \text{\bf Con}_{\text{\bf m}}^{\text{\bf spec}}$}\\ \hline\hline  
indicator  & description & elements \\ \hline
$\#$       & number of vortex cells & 1,2 \\ \hline
             &                        & {\bf N} (normal), \\
{\bf Con}  & configuration          & {\bf A} (anomalous), \\
             &                        & {\bf T} (twin-cell) \\
             &                        & {\bf M} (modulated rotating wave), \\ \hline
             &                        & {\bf z-osci} (axially oscillating), \\
             &                        & {\bf $\theta$-osci} (azimuthally oscillating), \\            
{\bf spec} & specification          & {\bf rot} (rotating) \\
             &                        & {\bf c} (compressed) \\
             &                        & {\bf *} (symmetry related) \\ \hline
{\bf m}    & stimulated modes       & 1, 2 \\ \hline
\end{tabular}
\end{center}
\caption{{Flow state nomenclature and abbreviations.}
}  
\label{tab:names}
\end{table}

\begin{figure}[]
\begin{center}  
\includegraphics[width=1.0\linewidth]{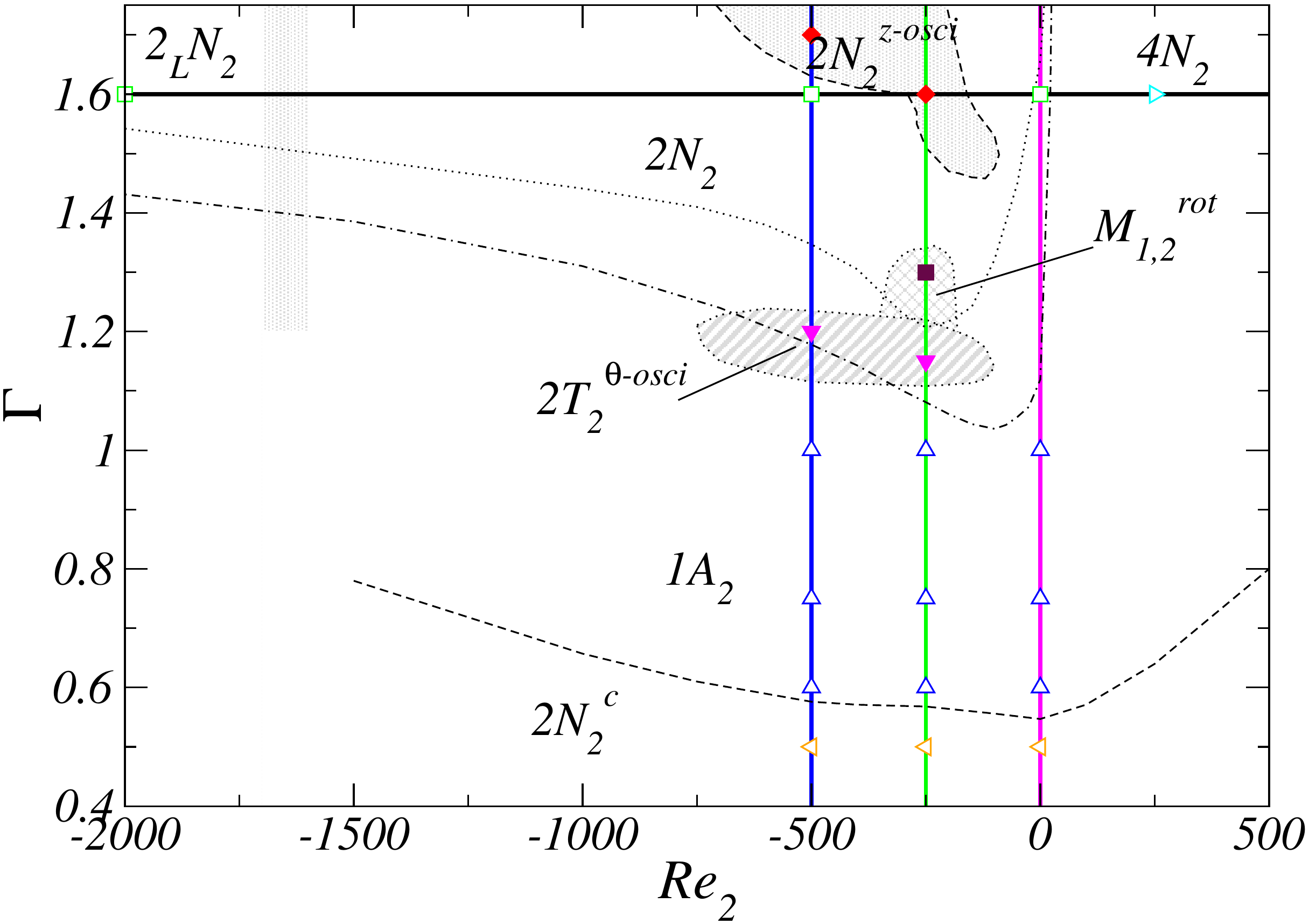}
\end{center}
\caption{{\bf Cases of simulations and analysis carried out in the parameter 
space ($\Gamma,Re_2$).} We focus on the variation in the Reynolds number of
the outer cylinder, $-2000 \leqslant Re_2 \leqslant 500$ for fixed aspect 
ratio $\Gamma=1.6$, or on variations in the aspect ratio for fixed 
$Re_2 = -500,-250$ and $0$, as indicated by the straight and horizontal 
lines, respectively. Different symbols highlight the parameters for which 
the flow states are studied in greater detail. The dashed and dotted lines 
specify the parameter values for which only a qualitative analysis of the
flow states is carried out in terms of their existence and stability. 
Within the region between the dotted and dashed-dotted curves, one-cell 
and two-cell flow states exist and are bistable.
}  
\label{fig:phasespace}
\end{figure}

\paragraph*{Parameter space and quantities.} 
Figure~\ref{fig:phasespace} provides an overview of the structure of the 
parameter space ($\Gamma,Re_2$) investigated in this paper. Simulations for
the parameters specified by the solid horizontal and the vertical lines are 
carried out and different symbols highlight the parameter values for 
which flow states are studied in great detail. 
For the dashed and dotted lines, the parameters are chosen according to
the steps $\Delta Re_2=50$  and $\Delta\Gamma=0.02$.

As a global measure to characterize the flow, we use the modal kinetic energy
defined as
\begin{eqnarray}
E_{kin} = \sum_{m} E_m = \int_0^{2\pi} \int_{-\Gamma/2}^{\Gamma/2}
\int_{r_i}^{r_o} {\bf u}_m {\bf u}^*_m r \textrm{d}r \textrm{d}z
\textrm{d}\theta, 
\end{eqnarray}
where {\bf u}$_m$ ({\bf u}$^*_m$) is the $m$-th (complex conjugate) Fourier 
mode of the velocity field, $E_{kin}$ is constant (non-constant) for a 
steady (an unsteady) solution. For a diagnostic purpose, we consider the 
time-averaged (over one period) quantity, 
$\overline{E}_{kin}= \int_{0}^{T} E_{kin} \textrm{d}t$. In addition to
the global measure, we also use the azimuthal vorticity on the inner 
cylinder at two points symmetrically displaced about the mid-plane, 
$\eta_\pm=(r_i,0,\pm\Gamma/4,t)$, as a local measure to characterize the
flow states (see {\bf Methods}). 
The unsteady, {\em oscillating} flow states in the axial and the azimuthal
directions are key dynamical states of the underlying TCS system with a
complex fluid. In order to obtain the axial and/or azimuthal frequencies,
we first conduct visualization of the full flow state
to decide if the flow structure is oscillating in the axial or
the azimuthal direction, or even rotating as a whole. In the case of
oscillating states (axial or azimuthal), the full flow fields are
identical after a period time $\tau$. The inverse of this period time
defines the frequency $\omega$. We then calculate the power spectral
densities (PSDs) of the global quantity $E_{kin}$ as well the local
quantities $\eta_\pm$, taking into account the system symmetries. These
two steps, together with the knowledge of the spatiotemporal behavior of
the flow structure, give the frequency of the flow state.
A more detailed description of our classication scheme for axially
oscillating, azimuthally oscillating, or rotating flow states is
presented in Supplementary Materials.

\begin{figure}[]
\begin{center}
\includegraphics[width=0.8\linewidth]{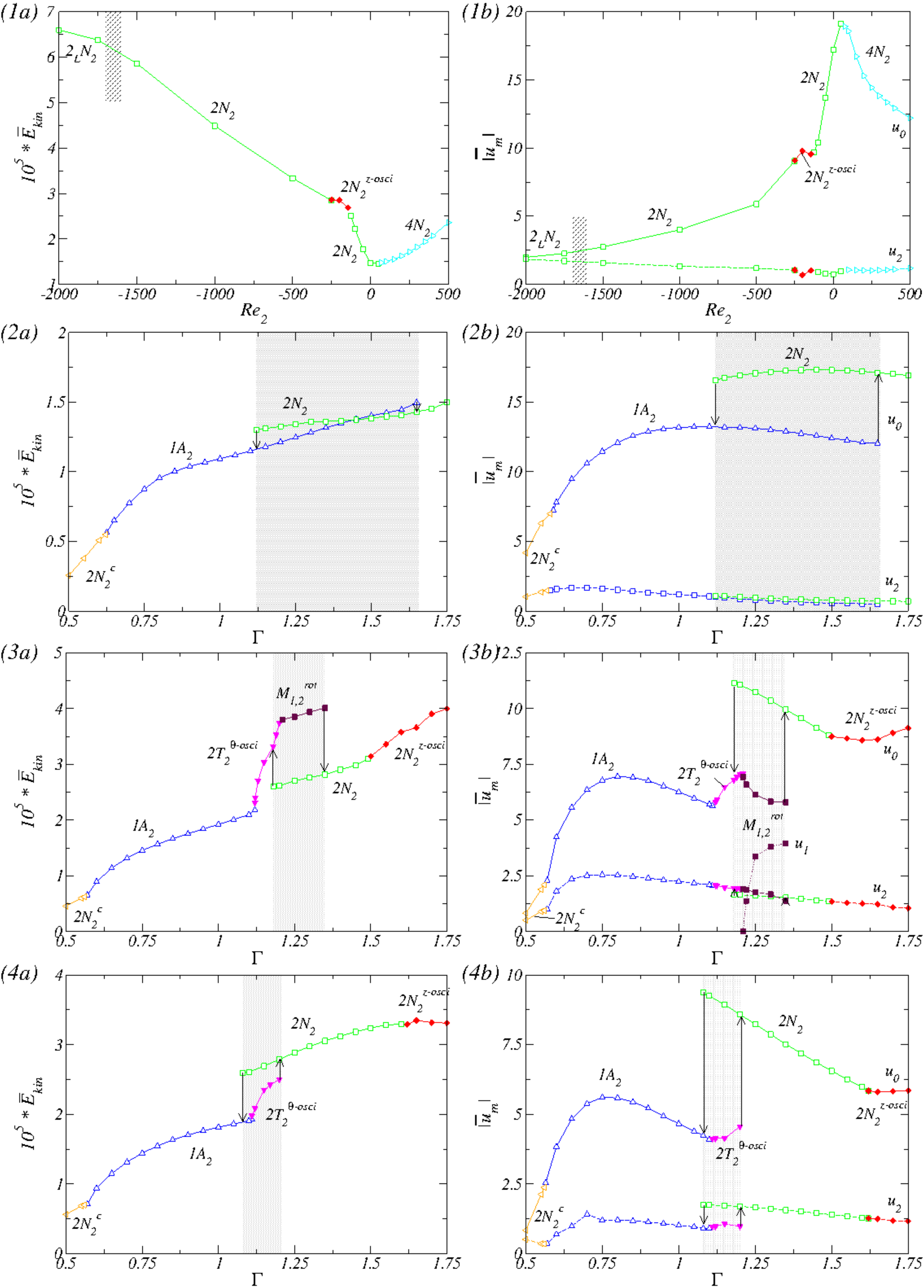}
\end{center}
\caption{{\bf Bifurcation scenarios.} 
Bifurcation scenarios with $(1)$ the Reynolds number $Re_2$ of the outer cylinder
at fixed $\Gamma=1.6$ and bifurcation with the aspect ratio $\Gamma$ for $(2)$ $Re_2=0$,
$(3)$ $Re_2=-250$ and $(4)$ $Re_2=-500$, respectively.
Shown are $(a)$ the total (time-averaged for oscillatory 
flows) modal kinetic energy $\overline{E}_{kin}$ and $(b)$ the corresponding
dominant (time-averaged) amplitudes $|\overline{u}_m|$ of the radial 
velocity field at mid-gap contributed by the axisymmetric mode [$u_0$, 
solid line in $(b)$] and the $m=2$ mode [$u_2$, dashed line in $(b)$].
Full (empty) symbols represent the time-dependent (time-independent) flows. 
Different flow structures are labeled. The highlighted gray areas indicate
the region of coexistence of distinct flow states: 
$Re_2=0: 1.119\leqslant\Gamma\leqslant1.657$;
$Re_2=-250: 1.18\leqslant\Gamma\leqslant1.34$;
$Re_2=-500: 1.108\leqslant\Gamma\leqslant1.21$.
}
\label{fig:bif_scenarios}
\end{figure}

\paragraph*{Bifurcation with $Re_2$.} 
To detect and understand the emergence of novel flow patterns in ferrofluidic
TCS with a small aspect ratio, we start from a moderate value 
so that the conventional two-cell flow state occurs~\cite{BeMu:1981,LMWP:1984,
PBE:1992,ADML2012}. To be concrete, we set $\Gamma=1.6$ and take $Re_2$ 
as a bifurcation parameter. Figure~\ref{fig:bif_scenarios}$(1)$ shows the 
total modal kinetic energy $\overline{E}_{kin}$ $(a)$ together with the axisymmetric 
[$u_0$, the solid line in $(b)$] and two-fold symmetric [$u_2$, dashed line 
in $(b)$] mode amplitudes versus $Re_2$. Note that the two-fold symmetric 
mode is intrinsically stimulated when a transverse magnetic field is 
present~\cite{SACHALML2010,ReOd:2011,ALD:2012,ALD:2013}.

\begin{figure}[]
\begin{center}
\includegraphics[width=0.7\linewidth]{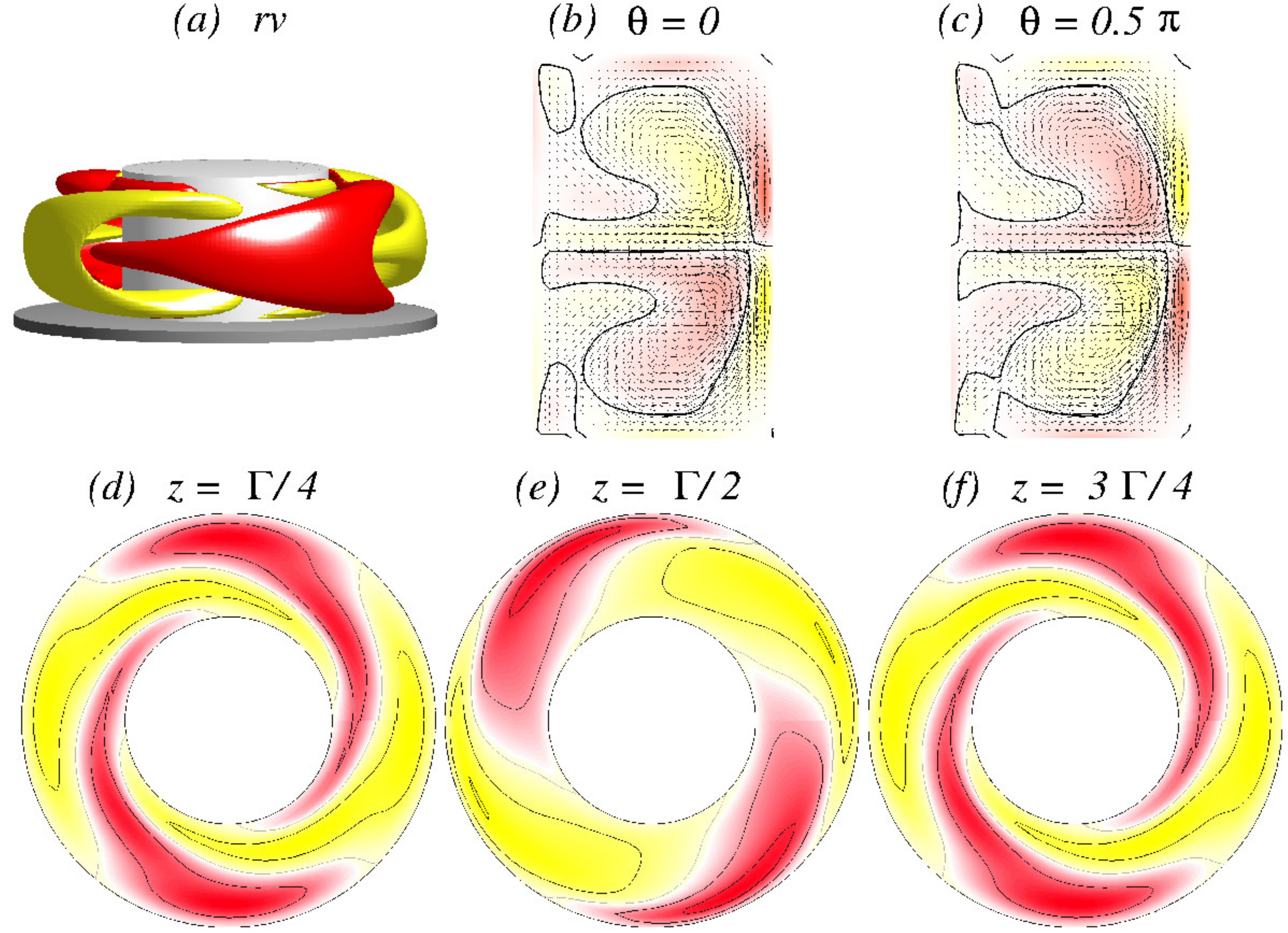}
\end{center}
\caption{ {\bf Flow visualization of the $2N_2$ state.}
Flow visualization of $2N_2$ for $\Gamma=1.6$ and $Re_2=0$: $(a)$ isosurface 
of $rv$ (isolevel shown at $rv=\pm5$) and the corresponding vector plot 
$[u(r,z),w(r,z)]$ of the radial and axial velocity components (including 
the azimuthal vorticity $\eta(r,\theta)$) for $(b)$ $\theta=0$ and $(c)$ 
$\theta=\pi/2$. (d-f) The azimuthal velocity $v(r,\theta)$ in three different 
planes: $z=\Gamma/4$, $z=\Gamma/2$, and $z=3\Gamma/4$, respectively. The same
legends are used for visualizing all the time independent flows in the paper.}  
\label{fig:G1_6_R20_iso_rv_r-z_r-phi}
\end{figure}

\subparagraph*{The case of $Re_2=0$.}
When the outer cylinder is at rest ($Re_2=0$), a two-cell flow state $2N_2$ 
is developed, as shown in Fig.~\ref{fig:G1_6_R20_iso_rv_r-z_r-phi}. Due
to the two-fold symmetry, this state differs little from the classical 
two-cell state in the absence of the magnetic field. The state
possesses both the $R_\pi^H$ and $K^H_z$ symmetries~\cite{ALD:2013} (see {\bf Methods}).
The isosurface plot for $rv=\pm5$ and the cross-sections in the $(r,z)$ 
plane indicate the two-fold rotational symmetry $R_\pi^H$ of the $2N_2$ state:
$\eta(r,\theta=0,z) = -\eta(r,\theta=\pi/2,z)$, whereas the horizontal cuts 
exhibit the $K^H_z$ invariance.

\begin{figure}[]
\begin{center}
\includegraphics[width=0.7\linewidth]{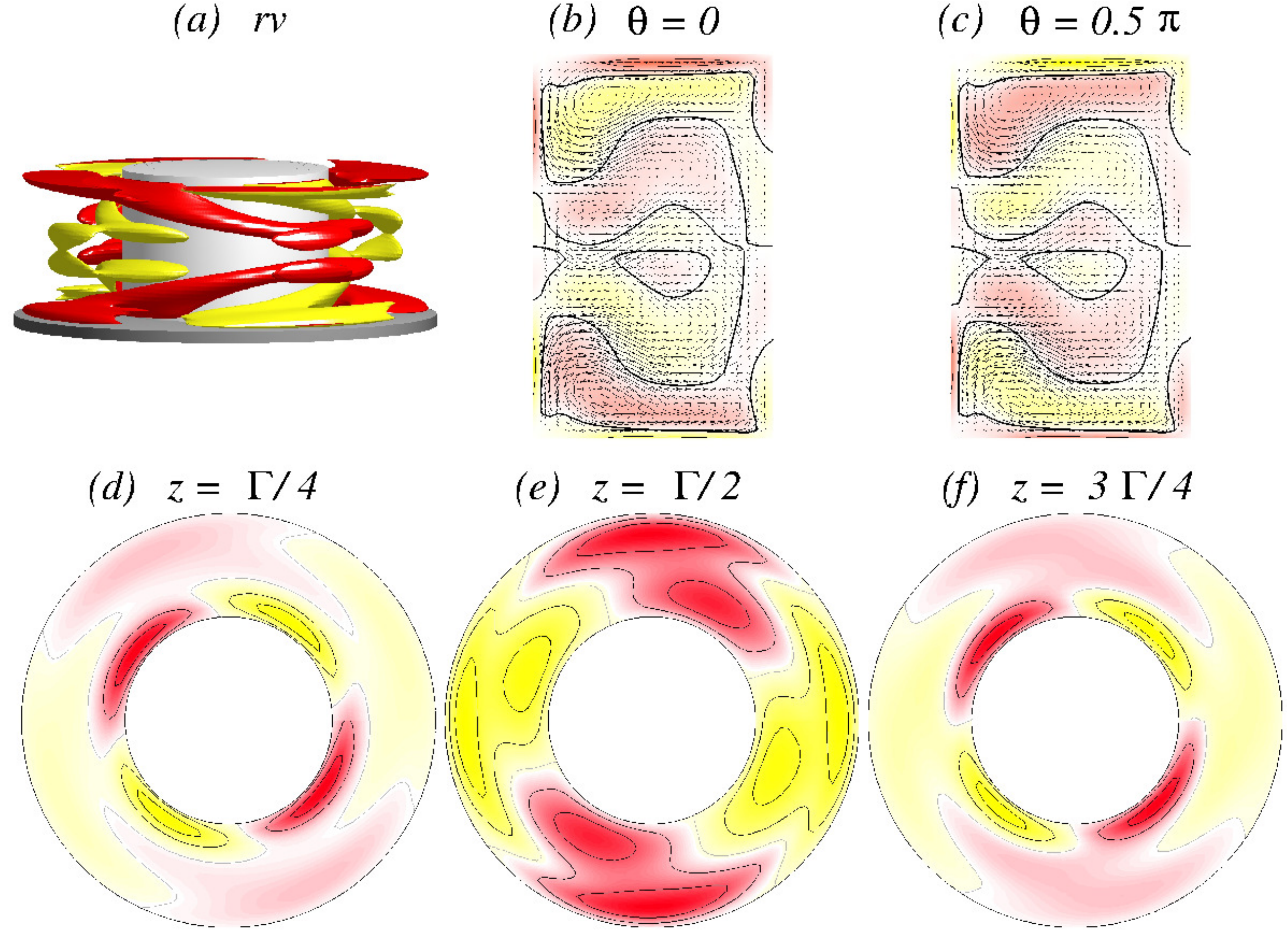}
\end{center}
\caption{{\bf Visualization of flow state $4N_2$.}
The $4N_2$ flow state for $\Gamma=1.6$ and $Re_2=250$. The legends 
are the same as in Fig.~\ref{fig:G1_6_R20_iso_rv_r-z_r-phi}. The 
isosurface for $rv=\pm7$ is shown in (a).}
\label{fig:G1_6_R2250_iso_rv_r-z_r-phi}
\end{figure}

\subparagraph*{The case of $Re_2>0$.}
Increasing the value of $Re_2$ from zero so that the cylinders rotate in 
the same direction, the flow starts to develop two additional symmetrical
vortex cells. These cells appear for $Re_2 \approx 27$, which are 
initially located in the corners near the inner cylinder and the lids.
As a result, a four-cell flow state emerges, denoted as $4N_2$, which has
the same symmetries ($K^H_z$, $R_\pi^H$) as the two-cell state $2N_2$ that it 
emerges from. An example of $4N_2$ for co-rotating cylinders at exactly 
the same speed is presented in Fig.~\ref{fig:G1_6_R2250_iso_rv_r-z_r-phi}.
As $Re_2$ is increased, the original two-cells are pulled closer 
towards the outer cylinder, giving more space for the two additional cells 
that extend into the interior of the bulk. This effect becomes continuously 
stronger for larger values of $Re_2$. For the parameter range investigated, 
i.e., $Re_2 \leqslant 500$, the flow state remains qualitatively the same as 
that shown in Fig.~\ref{fig:G1_6_R2250_iso_rv_r-z_r-phi}. Increasing $Re_2$ 
also leads the kinetic energy $\overline{E}_{kin}$ to increase continuously. 
During this process the contribution to the energy from the dominant 
axisymmetric $m=0$ mode ($u_0$) decreases while that from the $m=2$ 
mode ($u_2$) increases slightly, as shown in 
Fig.~\ref{fig:bif_scenarios}$(1)$.

\subparagraph*{The case of $Re_2<0$.}
As $Re_2$ is decreased from zero, the two cylinders become counter-rotating.
Initially the flow state $2N_2$ remains unchanged. As $Re_2$ is decreased 
through a critical value of about $-180$, the state $2N_2$ loses its stability 
via a supercritical symmetry-breaking Hopf bifurcation at which the mid-plane 
reflection symmetry $K^H_z$, together with a reversal of the magnetic 
field (cf., Eq.~\eqref{EQ:symmetry} and Ref.~\cite{ALD:2012} and {\bf Methods}), 
is broken and is replaced by a spatial temporal symmetry $S^H$ consisting of the 
mid-plane reflection $K^H_z$ in combination with a half-period time 
evolution $\Phi_{\tau/2}$. The physical manifestation of this symmetry
breaking phenomenon can be seen by noting that the two vortex cells now 
{\em oscillate axially} about the mid-plane. However, the new flow state 
$2N_2^\text{z-osci}$ is {\em not} a rotating state (which is the most typical 
case in TCS when the flow becomes time-dependent \cite{DiSw:1985,ALS:1986,Nag:1988}).

\begin{figure}[]
\begin{center}  
\includegraphics[width=0.7\linewidth]{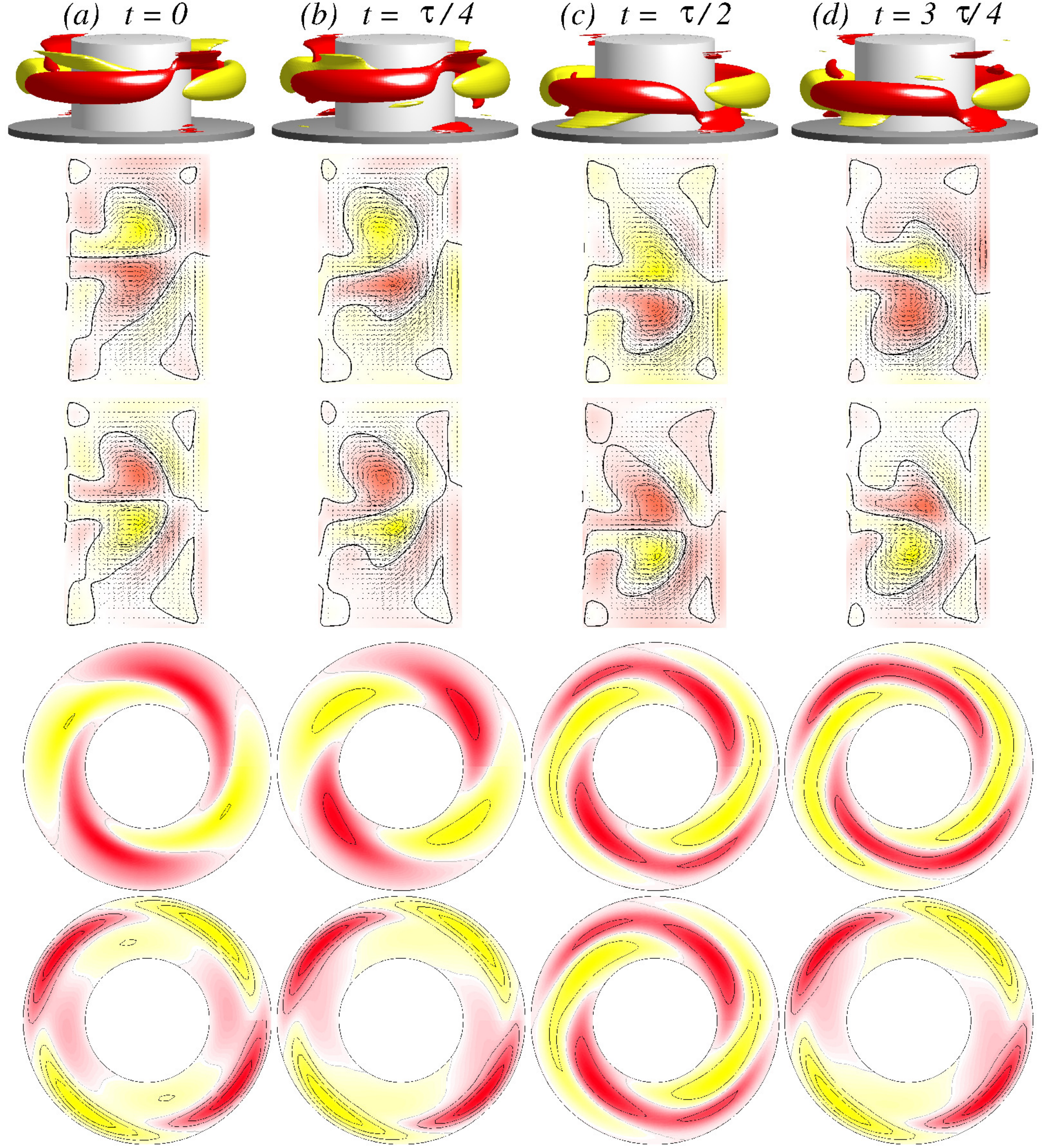}
\end{center}
\caption{ {\bf Visualization of the axially oscillating flow state 
$2N_2^\text{z-osci}$.} The first row shows, for $\Gamma = 1.6$ and 
$Re_2=-250$, the isosurfaces of $rv$ (isolevel shown at $rv=\pm15$) over 
one axially oscillating period ($\tau_z \approx 0.1635$) at instants of 
time $t$ as indicated. The second and third rows show the corresponding
vector plots $[u(r,z),w(r,z)]$ of the radial and axial velocity components
in the planes defined by $\theta=0$ and $\theta=\pi/2$, respectively,
where the color-coded azimuthal vorticity field $\eta$ is also shown. 
The fourth and fifth rows represent the azimuthal velocity $v(r,\theta)$ 
in the axial planes $z=\Gamma/4$ and $z=\Gamma/2$, respectively. Red (dark 
gray) and yellow (light gray) colors correspond to positive and negative 
values, respectively, with zero specified as white. See also movie file  
movieA1.avi, movieA2.avi, movieA3.avi and movieA4.avi
in Supplementary Materials (SMs) [The same axially oscillating
flow state $2N_2^\text{z-osci}$ for different parameters is preented in SMs:
movieE1.avi, movieE2.avi and movieE3.avi]. The same legends for
flow visualization are used for all subsequent unsteady flows.}
\label{fig:G1_6_R2-250_iso_rv_r-z}
\end{figure}

In order to get more insight the flow dynamics Fig.~\ref{fig:G1_6_R2-250_iso_rv_r-z}
presents four snapshots of the axially oscillating flow state $2N_2^\text{z-osci}$
[see also SMs: movieA1.avi movieA2.avi, movieA3.avi and movieA4.avi].
Shown are the angular 
momentum $rv$, vertical cross-section plots of $\eta(r,\theta=0[\pi/2],z)$, 
and horizontal cross-section plots of $v(r,\theta,z=1/4[1/2]\Gamma)$ over 
one period ($\tau_z \approx 0.1635$) illustrating the axial oscillation of 
the vortex cells. The figure also demonstrates the half-period flip 
symmetry $S^H$, where $K^H_z(2N_2^\text{z-osci}(t)) 
= 2N_2^\text{z-osci}(t+\tau_z/2)$. The effect of $S^H$ on the velocity field is
\begin{eqnarray}
\nonumber
S^H(u,v,w,H)(r,\theta,z,t) = (u,v,-w,-H)(r,\theta,-z,t+\tau_z/2).
\end{eqnarray}

\begin{figure}[]
\begin{center}
\includegraphics[width=1.0\linewidth]{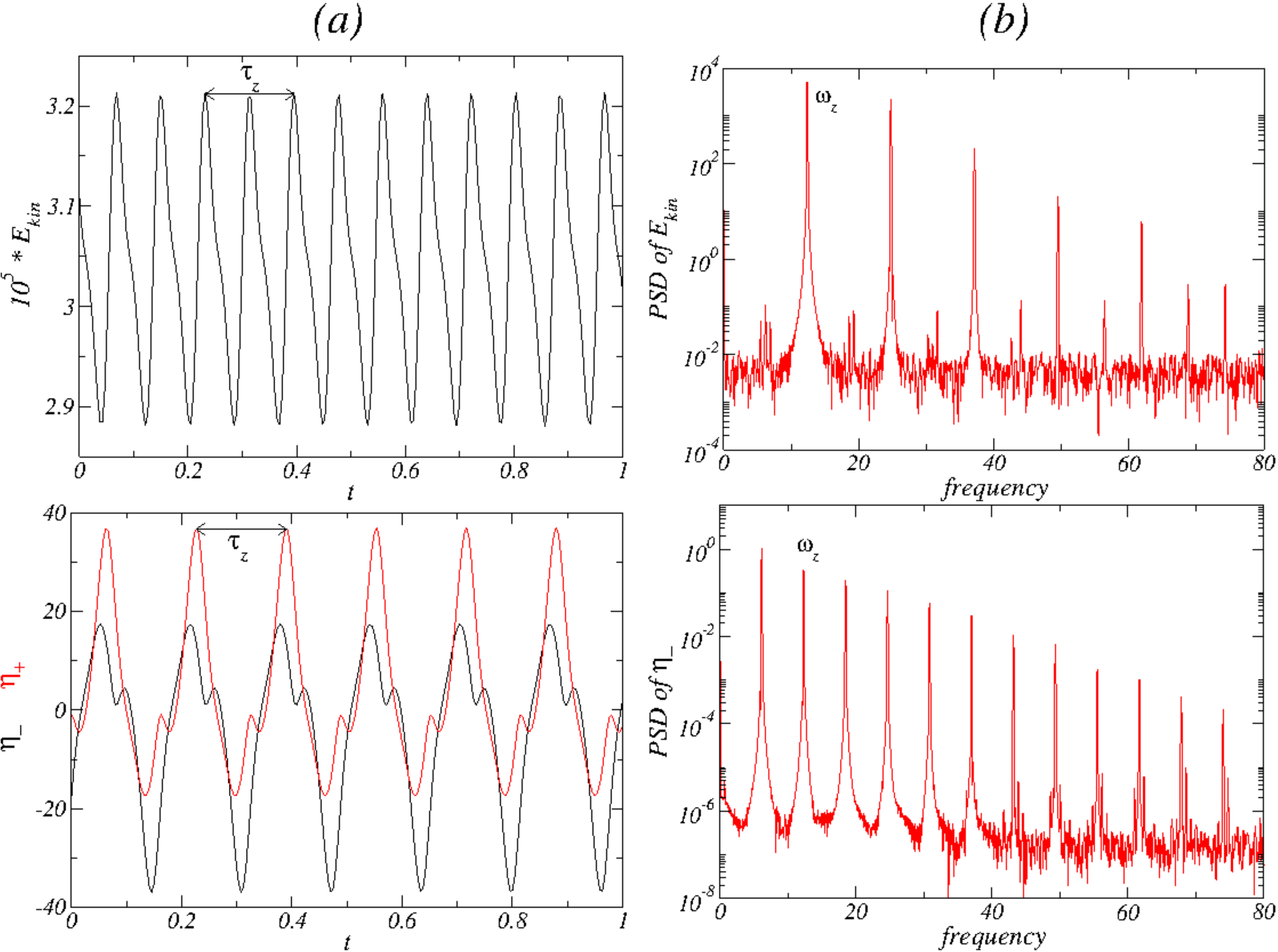}
\end{center}
\caption{{\bf Time series and PSD for the axially oscillating flow state 
$2N_2^\text{z-osci}$.} For $\Gamma = 1.6$ and $Re_2=-250$, $(a)$ time 
series of $E_{kin}$, $\eta_+$ [red (gray)], $\eta_-$ (black). $(b)$ The 
corresponding power spectral densities (PSDs) of the $2N_2^\text{z-osci}$ 
state. The period of axial oscillation is $\tau_z \approx 0.1635$ with
the corresponding frequency $\omega_z \approx 12.232$. Note that the PSD of
the local quantity $\eta_-$ has a peak at about half of this frequency,
indicating the half-period flip symmetry $S^H$ of the solution.}      
\label{fig:time_PSD_G1_6_R2-250_2N2osci}
\end{figure}

Topologically speaking the axially oscillating flow state $2N_2^\text{z-osci}$ is a limit 
cycle solution oscillating with the frequency $\omega_z$ in the axial direction
(cf., PSDs in Fig.~\ref{fig:time_PSD_G1_6_R2-250_2N2osci}). The state is 
thus qualitatively equivalent to the axially oscillating flow state in the 
classical TCS, which was first detected by 
Buzug {\em et al.}~\cite{BSP:1992}. The difference is that, in our 
ferrofluidic TCS, there is a finite contribution from the $m=2$ modes.
The limit cycle characteristic of $2N_2^\text{z-osci}$ results in closed curves
in the phase-space plot [see also SMs Fig.~9].

Figure \ref{fig:time_PSD_G1_6_R2-250_2N2osci} shows the time series of 
the modal kinetic energy $E_{kin}$ and $\eta_\pm$ together with its 
corresponding power spectral densities (PSDs) for the $2N_2^\text{z-osci}$ 
state for $\Gamma = 1.6$ and $Re_2=-250$. Note that $\tau_z$ is twice the 
period of the time series of $E_{kin}$ 
[cf., Fig.~\ref{fig:time_PSD_G1_6_R2-250_2N2osci}$(a)$], due to the 
fact that the $2N_2^\text{z-osci}$ state is half-period flip invariant and 
so $E_{kin}(2N_2^\text{z-osci}(t))=E_{kin}(2N_2^\text{z-osci}(t+\tau_z/2))$,
whereas if $2N_2^\text{z-osci}$ is $\tau_z$ periodic, we have
$2N_2^\text{z-osci}(t)=2N_2^\text{osci}(t+\tau_z)$. The half-period flip 
symmetry is visible in the time series of $\eta_+$ and $\eta_-$.

Further decreasing $Re_2$, the flow loses its time dependence again 
(through a similar symmetry breaking Hopf bifurcation but in the reverse
direction). For $Re_2 \approx -280$, the steady $2N_2$ flow state
emerges again. Due to the stronger counter rotation of the two cylinders,
the vortex cells are slightly shifted towards the inner cylinder wall but 
remain qualitatively the same, as shown for the case $Re_2=0$ in 
Fig.~\ref{fig:G1_6_R20_iso_rv_r-z_r-phi}.

As $Re_2$ is decreased continuously, the solution remains 
topologically identical to a two-cell flow state. For $Re$ about -1680, the 
flow undergoes a smooth transition in which the vortex centers are 
pushed towards the inner cylinder and the cells become elongated in the
axial direction. A slightly elongated two-cell flow 
state $2_LN_2$ is shown in SMs in Fig.~1. 
Except for the small parameter regime in which $2N^\text{z-osci}_2$ 
exists, the kinetic energy $\overline{E}_{kin}$ increases continuously 
with $Re_2$ (cf., Fig.~\ref{fig:bif_scenarios}$(1)$), where the 
contribution $u_0$ from the dominant mode $m=0$ decreases but that from 
the $m=2$ mode, $u_2$, increases. This is confirmed by the fact
that stronger counter-rotation flows favor higher azimuthal modes.

\paragraph*{Bifurcation with the aspect ratio $\Gamma$.} 
We now fix the value of $Re_2$ (at 0, -250 and -500), and investigate 
the bifurcation of the flow state with the aspect ratio $\Gamma$, respectively.
Note that, in the classical TCS, there can be a transition in the 
flow between two-cell and one-cell states as $\Gamma$ is varied. We aim
to uncover the similarity and difference in the bifurcations in the
ferrofluidic TCS.

\begin{figure}[]
\begin{center}
\includegraphics[width=0.7\linewidth]{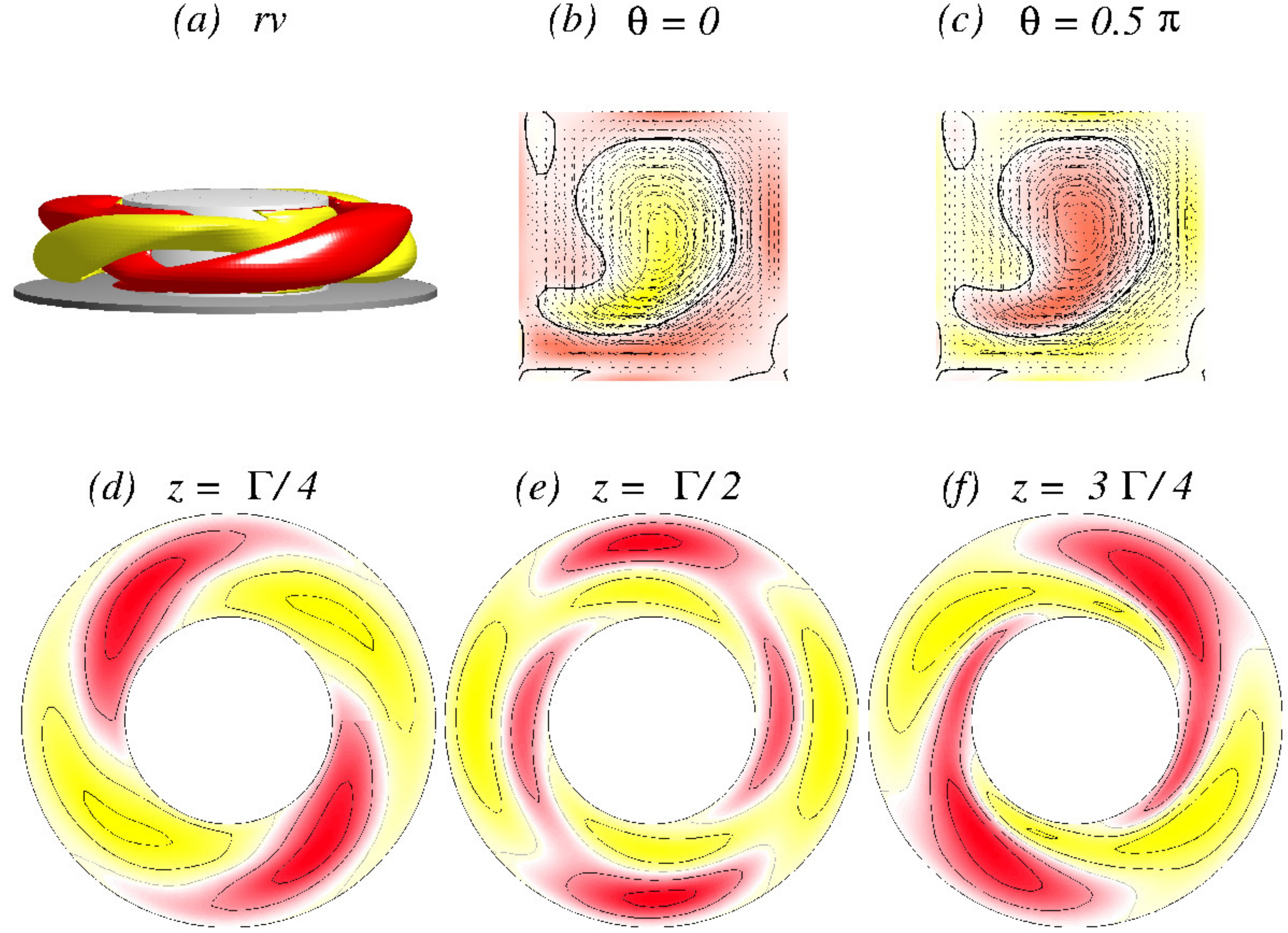}
\end{center}
\caption{{\bf Visualization of flow state $1A_2$} for $\Gamma=1.0$ and 
$Re_2=0$, where panel $(a)$ shows the isosurface for $rv=\pm7$. 
The legends are the same as in Fig.~\ref{fig:G1_6_R20_iso_rv_r-z_r-phi}.}
\label{fig:G1_0_R20_iso_rv_r-z_r-phi}
\end{figure}

\subparagraph*{The case of $Re_2=0$.}
Figure~\ref{fig:bif_scenarios}$(2)$ shows, for $Re_2=0$, the 
variation with $\Gamma$ of the modal kinetic energy $\overline{E}_{kin}$ 
and the dominant amplitude $|\overline{u}_m|$ associated with the flow.
Increasing $\Gamma$ from 1.6 (cf., Figs.~\ref{fig:bif_scenarios}$(1)$ 
and \ref{fig:G1_6_R20_iso_rv_r-z_r-phi}), the flow state $2N_2$ remains 
unchanged and stable until $\Gamma=1.75$. Decreasing $\Gamma$ from 1.6, 
the same state holds (cf., Fig.~\ref{fig:bif_scenarios}$(2)$) until 
when $\Gamma \approx 1.12$, where the $2N_2$ state loses its stability 
and becomes a transient. The final flow state has only {\em one} dominant 
vortex cell, i.e., the one-cell flow state $1A_2$. 
Figure~\ref{fig:G1_0_R20_iso_rv_r-z_r-phi} illustrates the $1A_2$ state
for $\Gamma=1.0$, where the single vortex nature is apparent. While 
$1A_2$ itself is not $K^H_z$ symmetric, there is a coexisting, symmetry 
possessing flow state $1A^*_2$ that can be obtained by applying the 
symmetry operation $K^H_z$: $1A^*_2=K^H_z 1A_2$. (See below for a 
detailed explanation). Representative isosurfaces of $rv$ are shown in
Fig.~\ref{fig:G1_0_R20_iso_rv_r-z_r-phi}), where a twisted vortex 
structure with a two-fold symmetry due to the magnetic field can be 
identified.

\begin{figure}[]
\begin{center}
\includegraphics[width=0.7\linewidth]{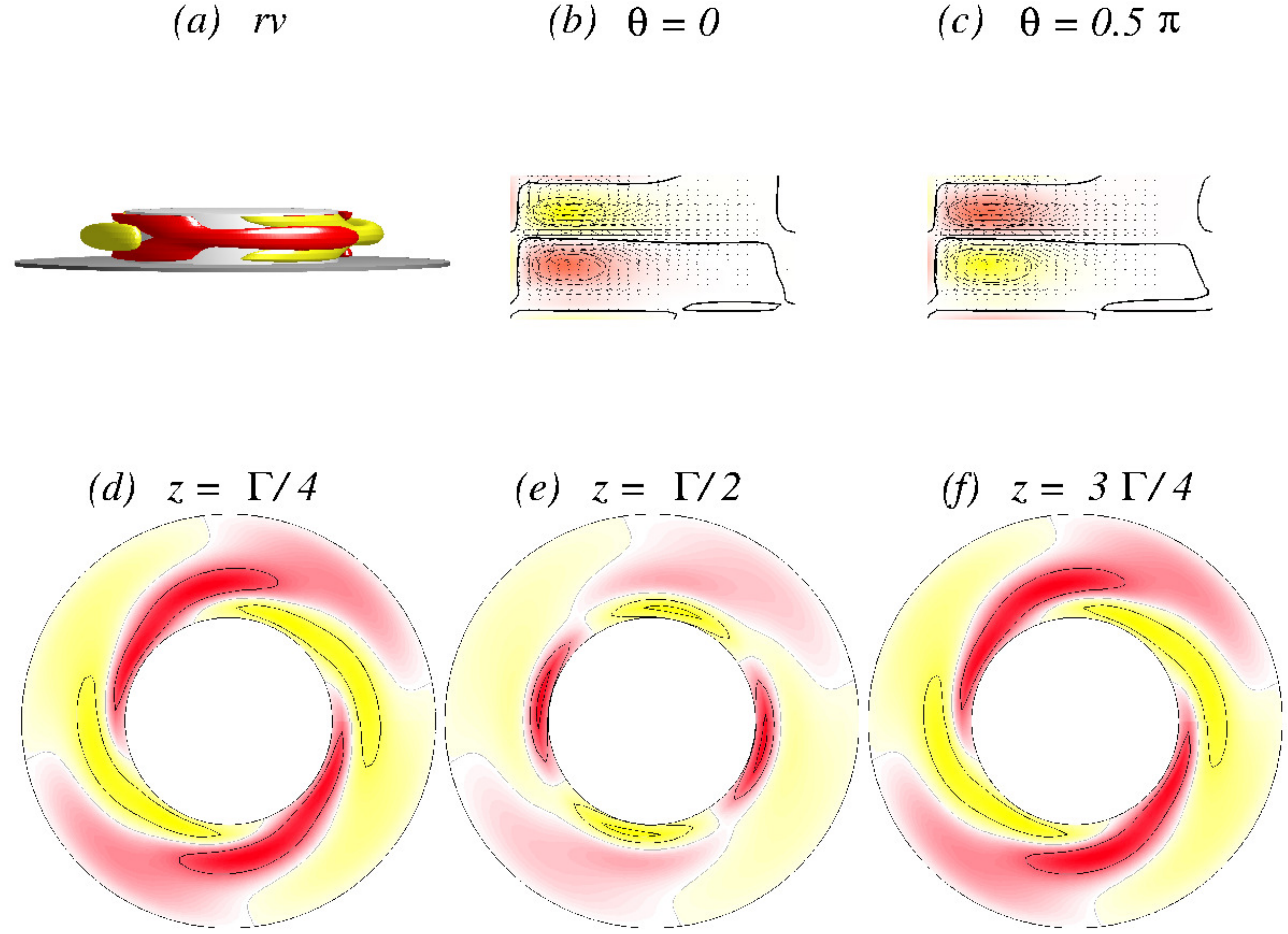}
\end{center}
\caption{{\bf Visualization of flow state $2N_2^\text{c}$} for
$\Gamma=0.5$ and $Re_2=0$, where $(a)$ shows the isosurface for $rv=\pm5$.
The legends are the same as in Fig. \ref{fig:G1_6_R20_iso_rv_r-z_r-phi}. 
Note that this flow state was first reported in the classical TCS by 
Pfister {\em et al.}~\cite{BSP:1992}, who described it as a two-cell 
state with two compressed vortices near the inner cylinder.}
\label{fig:G0_5_R20_iso_rv_r-z_r-phi}
\end{figure}

\begin{figure}[]
\begin{center}
\includegraphics[width=0.7\linewidth]{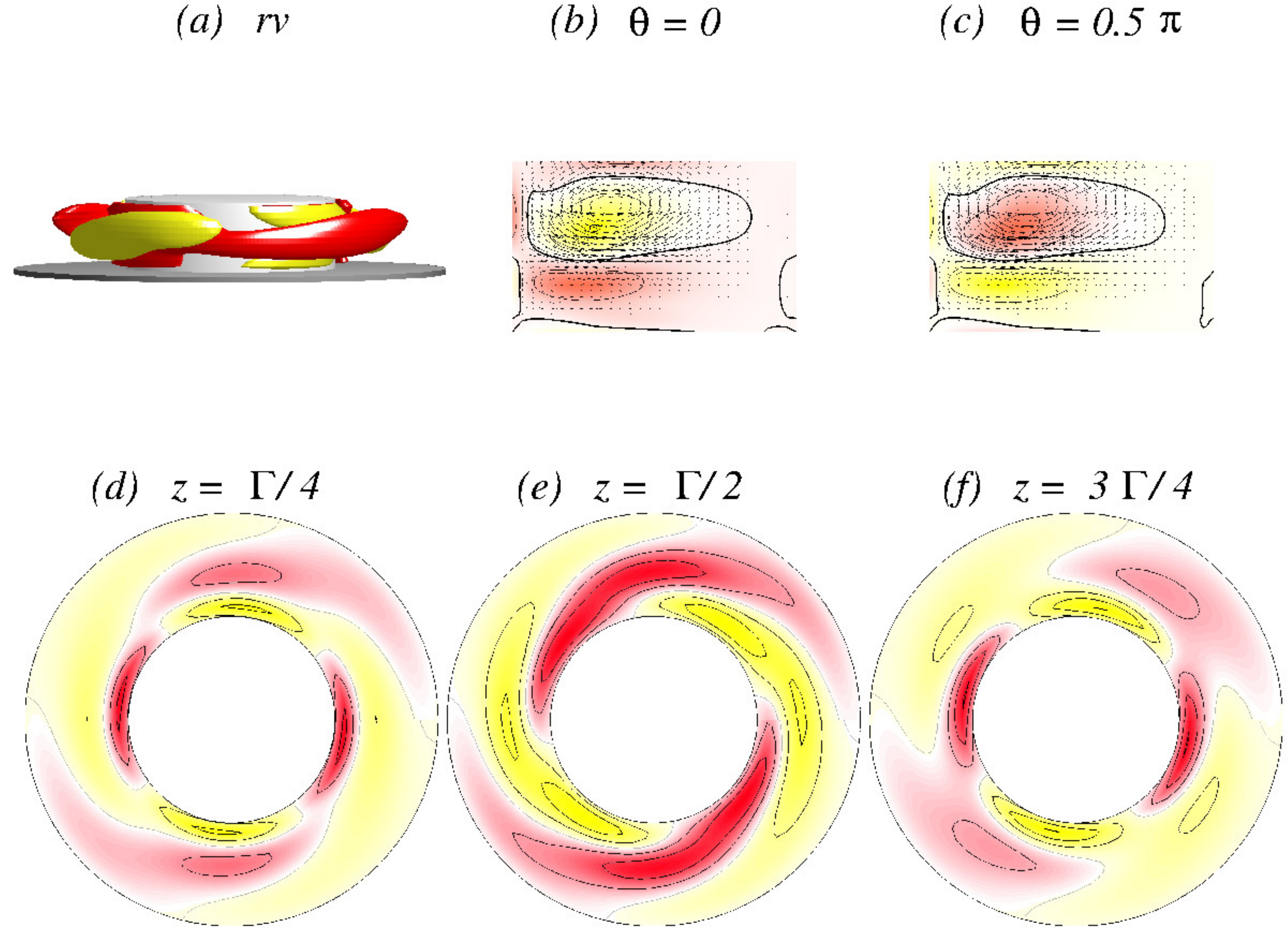}
\end{center}
\caption{{\bf Visualization of flow state $1A_2$} for $\Gamma=0.6$ and 
$Re_2=0$, where $(a)$ shows the isosurface for $rv=\pm5$. The legends are
the same as in Fig.~\ref{fig:G1_6_R20_iso_rv_r-z_r-phi}.}
\label{fig:G0_6_R20_iso_rv_r-z_r-phi}
\end{figure}

For clarity and simplicity we now describe the bifurcation sequence and
the evolution of the flow state as $\Gamma$ is increased from the relatively
small value of $0.5$. For $\Gamma \lesssim 0.57$, we find the two-cell 
flow state $2N_2^\text{c}$ (cf., Fig.~\ref{fig:G0_5_R20_iso_rv_r-z_r-phi}). 
Similar to the flow state $2N_2$, also $2N_2^\text{c}$ consists 
of two vortex cells but with the difference that the cells are compressed 
near the inner cylinder, which results in a wide (outer) region in which 
the annulus is essentially vortex free. Such states were first reported
by Pfister {\em et al.}~\cite{BSP:1992} for the classical TCS.
The flow state $2N_2^\text{c}$ for $\Gamma=0.5$, shown in 
Fig.~\ref{fig:G0_5_R20_iso_rv_r-z_r-phi} which differs from the classical one in only one 
aspect: a two-fold symmetry induced by the magnetic field~\cite{ALD:2013}.
Increasing $\Gamma$ further the state $2N_2^\text{c}$ undergoes a 
symmetry breaking pitchfork bifurcation at $\Gamma \approx 0.57$, where 
the $K_z^H$ symmetry is broken and one of the vortex cells grows at the 
cost of the other. As a result, two symmetry related one-cell flow states 
emerge: $1A_2$ and $1A_2^*=K_z^H A_2$. For convenience, from here on we 
denote them as a one-cell flow states $1A_2$ ($1A_2^*$), even when one 
of the vortex cells is slightly smaller than the other. The main dynamics 
is essentially dominated by the large one. That is, when 
$1A_2$ is mentioned, the coexistence of the $1A_2^*$ state is implied. An 
example for $1A_2$ after the symmetry breaking bifurcation is shown in 
Fig.~\ref{fig:G0_6_R20_iso_rv_r-z_r-phi} for $\Gamma=0.6$, where it can 
be seen that the top vortex cell has started to grow (from the bifurcation 
point) at the cost of the bottom cell. On the $rv$ isosurfaces,
the formation of the twisted vortex shape can be seen, as a result of 
the transverse magnetic field, which is typical for $1A_2$ 
(cf., Fig.~\ref{fig:G1_0_R20_iso_rv_r-z_r-phi}).

\begin{figure}[]
\begin{center}
\includegraphics[width=0.7\linewidth]{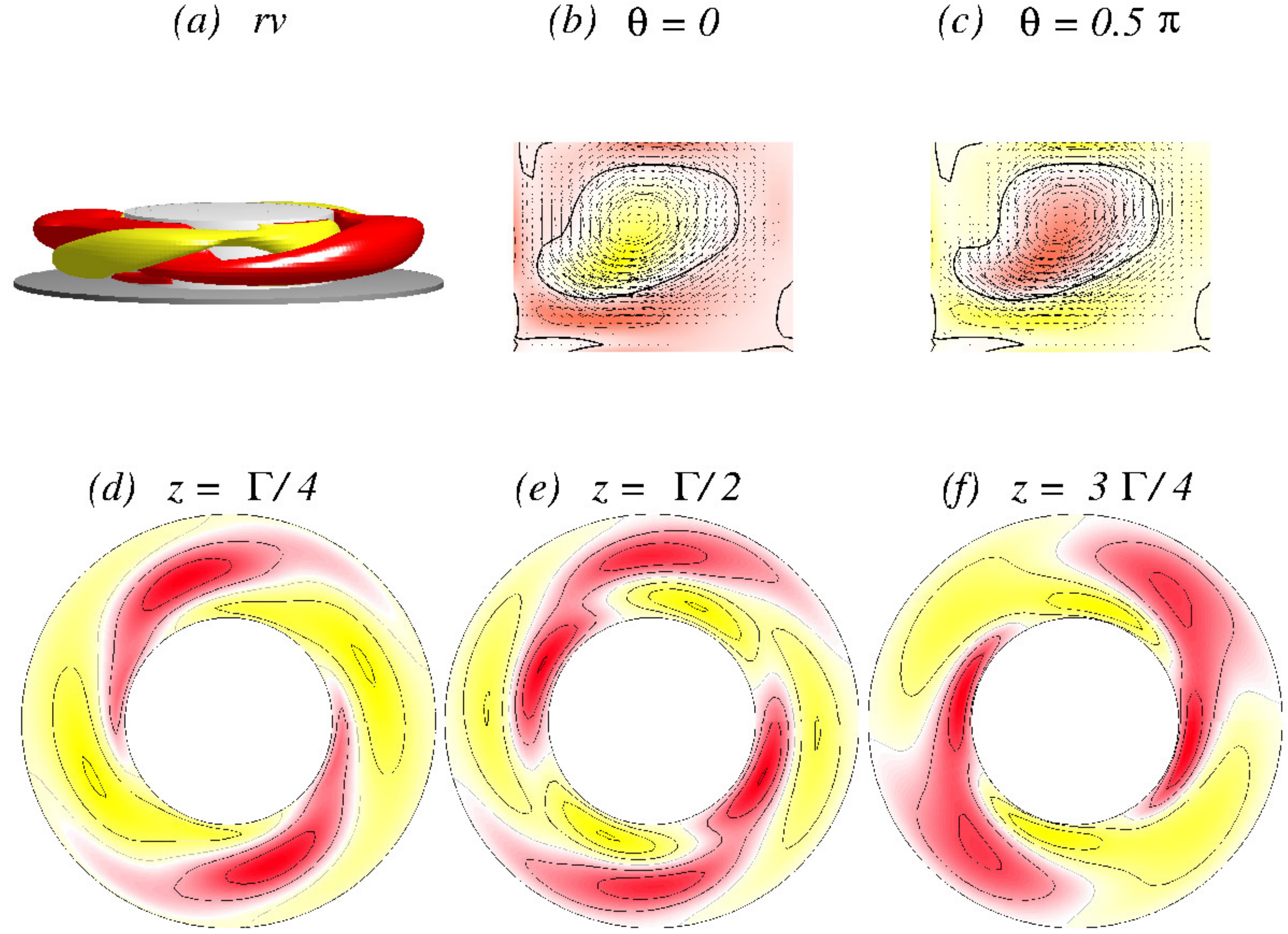}
\end{center}
\caption{{\bf Visualization of flow state $1A_2$} for $\Gamma=0.75$ and
$Re_2=0$, where $(a)$ shows the isosurface for $rv=\pm5$. Legends are the
same as in Fig.~\ref{fig:G1_6_R20_iso_rv_r-z_r-phi}.}
\label{fig:G0_75_R20_iso_rv_r-z_r-phi}
\end{figure}

As $\Gamma$ is increased further, one of the vortex cells continuously grows
in size to occupy more of the interior region of the bulk, while the other 
becomes increasingly compressed, which can be seen for the $1A_2$ state in 
Fig.~\ref{fig:G0_75_R20_iso_rv_r-z_r-phi} for $\Gamma=0.75$, where the 
dominance of one vortex cell is apparent. In fact, the 
flow state $1A_2$ exists in a wide range of $\Gamma$ values until it 
finally loses stability at $\Gamma \approx 1.66$ and becomes a transient 
to the two-cell flow state $2N_2$ that exists for even larger values of 
$\Gamma$ (e.g., even for $\Gamma=1.75$ - the largest value of the aspect
ratio studied in this paper). These results indicate that, in the parameter
interval $1.12 \lesssim \Gamma \lesssim 1.66$, there are two bistable coexisting 
flow states: a two-cell state $2N_2$ and a one-cell state $1A_2^{*}$. Both,
$2N_2$ and $1A_2^{*}$ are living on two different {\em not} connected 
solution branches.

\subparagraph*{The case of $Re_2=-250$.}
We now study the case of counter-rotating cylinders with $Re_2 < 0$. 
The corresponding bifurcation scenario is illustrated in Fig.~\ref{fig:bif_scenarios}$(3)$.
As for $Re_2=-250$ and $\Gamma=1.6$ (cf. Figs.~\ref{fig:bif_scenarios}$(1,2)$),
there is an unsteady, axially oscillating flow state $2N^\text{z-osci}_2$,
which serves as the baseline for our bifurcation analysis. Increasing 
$\Gamma$ from 1.6, this flow state is stable and continues to exist 
until for $\Gamma=1.75$, whereas if $\Gamma$ is decreased from 1.6, the
state loses its time dependence through a symmetry-breaking, backward 
Hopf bifurcation for $\Gamma \approx 1.48$, leading again to the flow 
state $2N_2$ (cf., Fig.~\ref{fig:G1_6_R20_iso_rv_r-z_r-phi}). This is
the same bifurcation scenario as described for constant $\Gamma$ while
varying $Re_2$ (cf., Fig.~\ref{fig:bif_scenarios}$(1)$). Upon 
further decrease in $\Gamma$, the $2N_2$ state remains stable until 
for $\Gamma \approx 1.08$ when it loses stability and simultaneously,
a single cell state $1A_2$ ($1A^*_2$) emerges (as described for $Re_2=0$).

Starting again with a small aspect ratio and increasing $\Gamma$, the
bifurcation sequence is the same as for $Re_2=0$. In particular, for 
$\Gamma=0.5$ the flow state $2N_2^\text{c}$ is present, which as $\Gamma$
is increased undergoes a symmetry breaking pitchfork bifurcation to the 
symmetry related one-cell flow state $1A_2$ ($1A^*_2$). The evolution of 
$1A_2$ as $\Gamma$ is increased toward unity is qualitatively the same as
for the case of $Re_2=0$ [see Fig.~2 in Supplementary Materials], with the 
small difference being that the vortices have slightly moved 
towards the inner cylinder due to the counter rotation.

\begin{figure}[]
\begin{center}  
\includegraphics[width=0.7\linewidth]{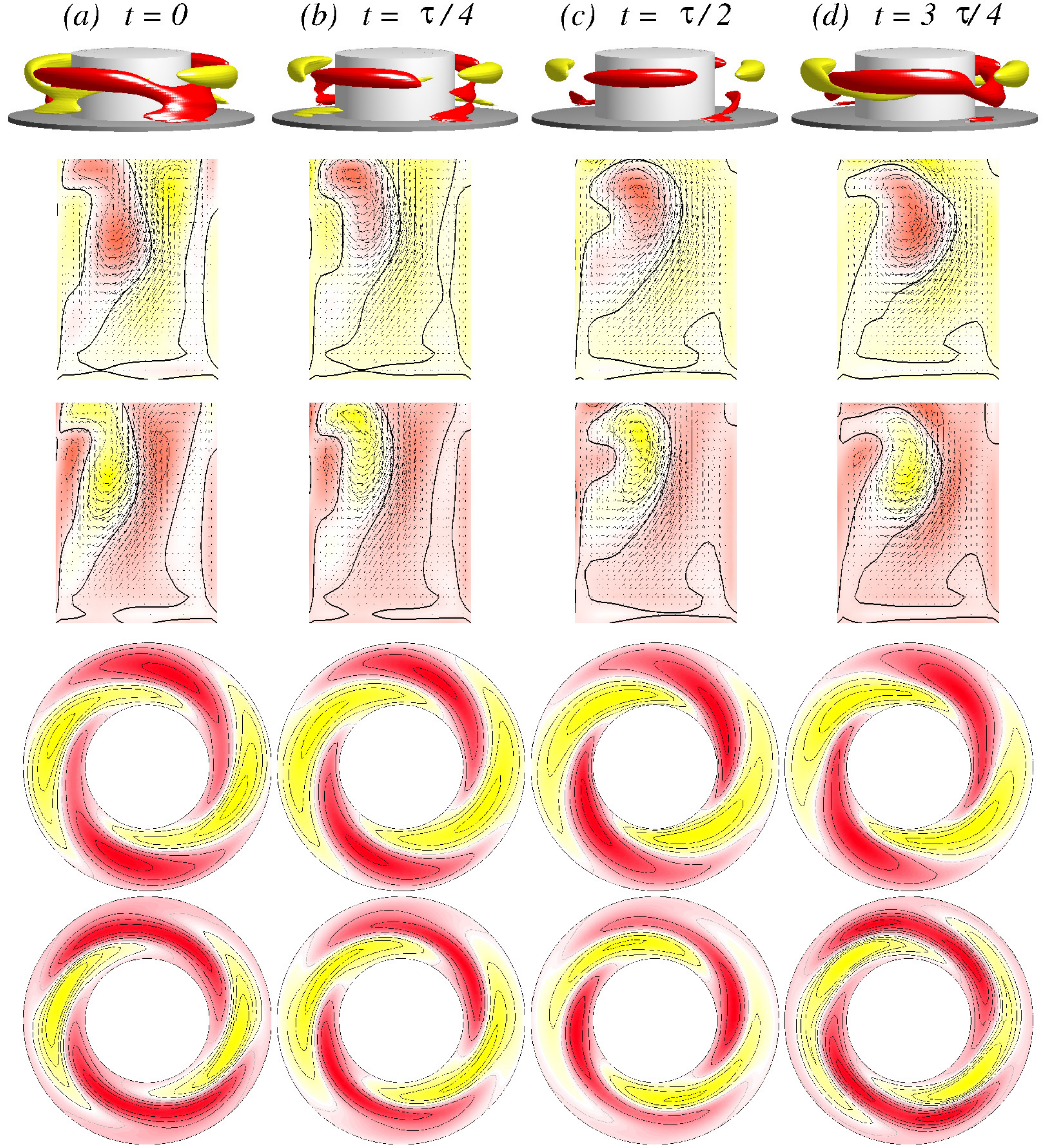}
\end{center}
\caption{ {\bf Visualization of the azimuthally oscillating twin-cell flow 
state $2T_2^{\theta \text{-osci}}$} for $\Gamma=1.15$, as in 
Fig.~\ref{fig:G1_6_R2-250_iso_rv_r-z}. The top row shows the isosurfaces 
for $rv=\pm25$. The oscillation period is $\tau_\theta \approx 0.0954$.
See movie files movieB1.avi, movieB2.avi, movieB3.avi and movieB4.avi 
and movieB4.avi in SMs.}
\label{fig:G1_15_R2-250_iso_rv_r-z}
\end{figure}

With further increase in $\Gamma$ the steady one-cell flow state $1A_2$ 
undergoes a Hopf bifurcation, at which an unsteady two-cell flow state
emerges with frequency $\omega_\theta$. For 
$\Gamma \approx 1.12$, the flow starts to oscillate in the {\em azimuthal} 
direction with the frequency $\omega_\theta$. A difference from the case
of axially oscillating flow $2N^\text{z-osci}_2$ (cf., 
Figs.~\ref{fig:G1_6_R2-250_iso_rv_r-z} and 
\ref{fig:time_PSD_G1_6_R2-250_2N2osci}) is that the new oscillating flow 
state $2T^{\theta \text{-osci}}_2$ has the characteristics of a 
{\em twin-cell} flow state, meaning that the vortex cells are arranged 
side by side and they both touch the top and the bottom lids, as shown in 
Fig.~\ref{fig:G1_15_R2-250_iso_rv_r-z}, instead of being on top of 
each other. Topologically this flow corresponds to a limit cycle 
solution,
similar to the $2N^\text{z-osci}_2$ state.

\begin{figure}[]
\begin{center}
\includegraphics[width=1.0\linewidth]{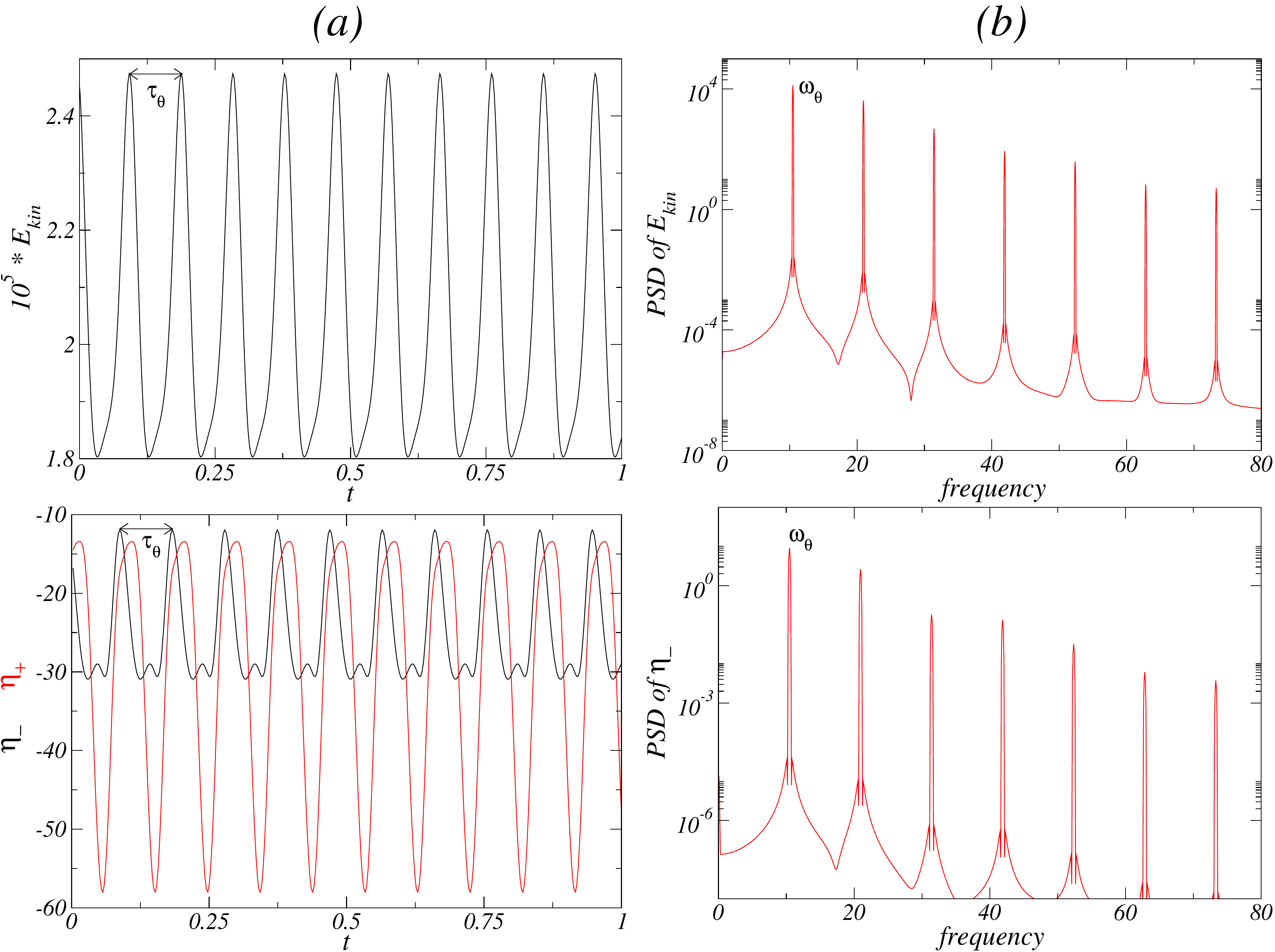}
\end{center}
\caption{ {\bf Time series and PSD of the azimuthally oscillating flow 
state $2T_2^{\theta \text{-osci}}$.} For $\Gamma = 1.15$ and $Re_2=-250$,
$(a)$ time series of $E_{kin}$, $\eta_+$ [red (gray)], $\eta_-$ (black), 
and $(b)$ the corresponding PSDs for $2T_2^{\theta \text{-osci}}$.
The period of the azimuthal oscillation is $\tau_\theta \approx 0.0954$ with
the corresponding frequency $\omega_\theta \approx 10.482$.}
\label{fig:time_PSD_G1_15_R2-250_2T2osci}
\end{figure}

Figure~\ref{fig:G1_15_R2-250_iso_rv_r-z} demonstrates the
$2T_2^{\theta \text{-osci}}$ flow state for $\Gamma=1.15$
[see also Supplementary Materials: movieB1.avi, movieB2.avi, movieB3.avi
and movieB4.avi]. 
There are four snapshots of the angular momentum $rv$, vertical 
cross-section plots of $\eta(r,\theta=0[\pi/2],z)$, and horizontal cross-section 
plots of $v(r,\theta,z=1/4[1/2]\Gamma)$ over one oscillating period 
$\tau_\theta \approx 0.0954$. Similar to the $2N^\text{z-osci}_2$ state,
$2T_2^{\theta \text{-osci}}$ oscillates but in the {\em azimuthal} 
direction. Moreover the $2T_2^{\theta \text{-osci}}$ state has no 
apparent symmetries (cf., Fig.~\ref{fig:G1_15_R2-250_iso_rv_r-z}), 
in particular no shift-reflect symmetry $S^H$ as $2N^\text{z-osci}_2$ has. 
The broken symmetry can also be seen in the corresponding time series 
of $\eta_+$ and $\eta_-$ in Fig.~\ref{fig:time_PSD_G1_15_R2-250_2T2osci}. 
Note that here the flow state $2T_2^{\theta \text{-osci},*}$ coexists, which 
evolves in the same way from the symmetry related flow state $1A_2^*$ 
(instead of $1A_2$) with increasing $\Gamma$. Analogous to the case of 
one-cell flow state, the symmetry related flow state 
$2T_2^{\theta \text{-osci},*}=K_z^H 2T_2^{\theta \text{-osci}}$ is
characterized by a reflection invariance at the mid-height.

\begin{figure}[]
\begin{center}  
\includegraphics[width=0.7\linewidth]{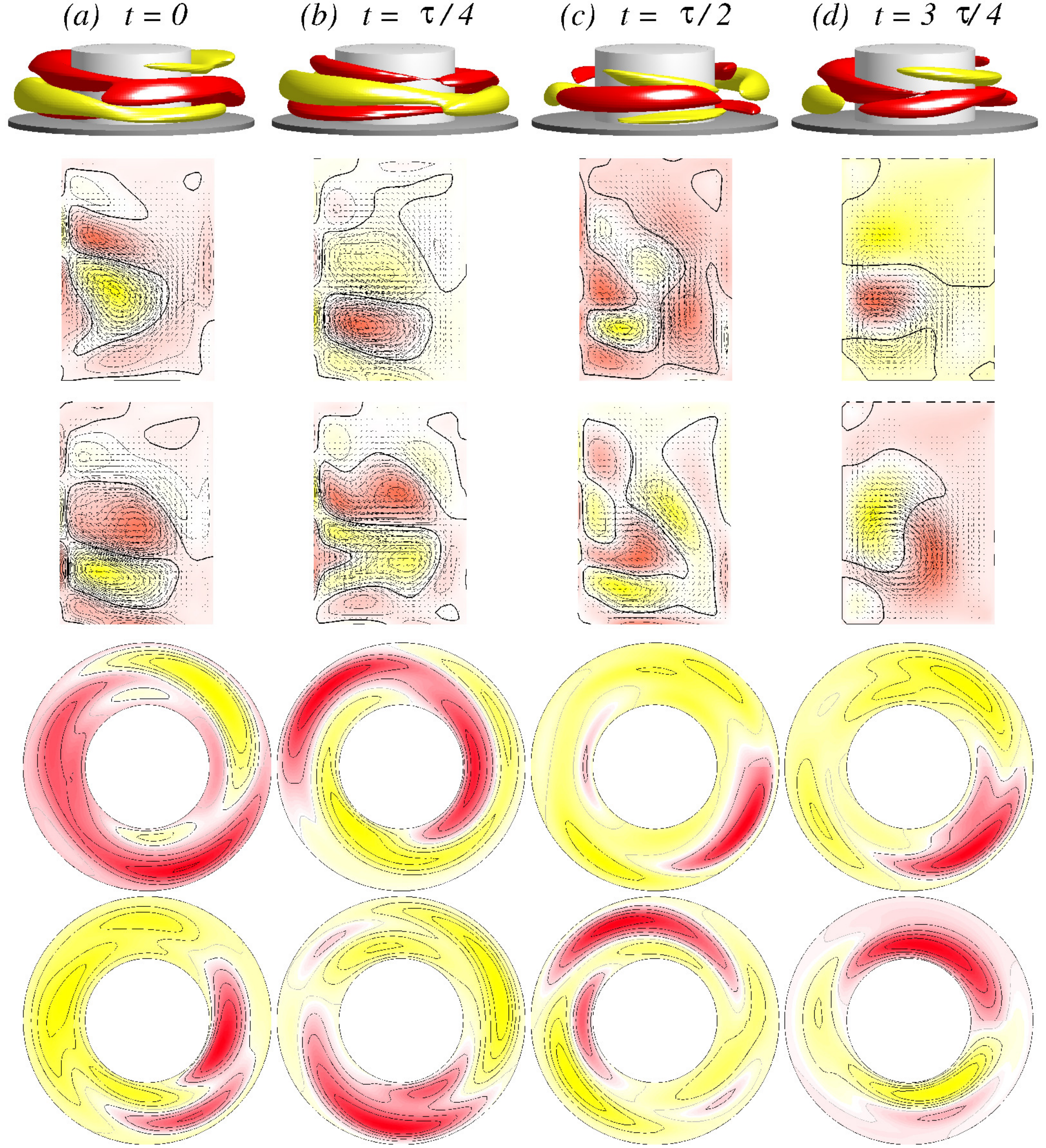}
\end{center}
\caption{ {\bf Visualization of rotating flow state $M_{1,2}^\text{rot}$}
for $\Gamma=1.3$ with the same legends as in 
Fig.~\ref{fig:G1_6_R2-250_iso_rv_r-z}. The top row shows the isosurfaces 
for $rv=\pm25$. The period of rotation is $\tau_\text{rot} \approx 0.7829$.
See movie file movieD1.avi and movieD2.avi in SMs. The vectorplots
in $(r,z)$ plane highlight very well the complexity of $M_{1,2}^\text{rot}$, in
particular the change from one-two-three-four cell states.
}
\label{fig:G1_3_R2-250_iso_rv_r-z}
\end{figure}

\begin{figure}[]
\begin{center}
\includegraphics[width=1.0\linewidth]{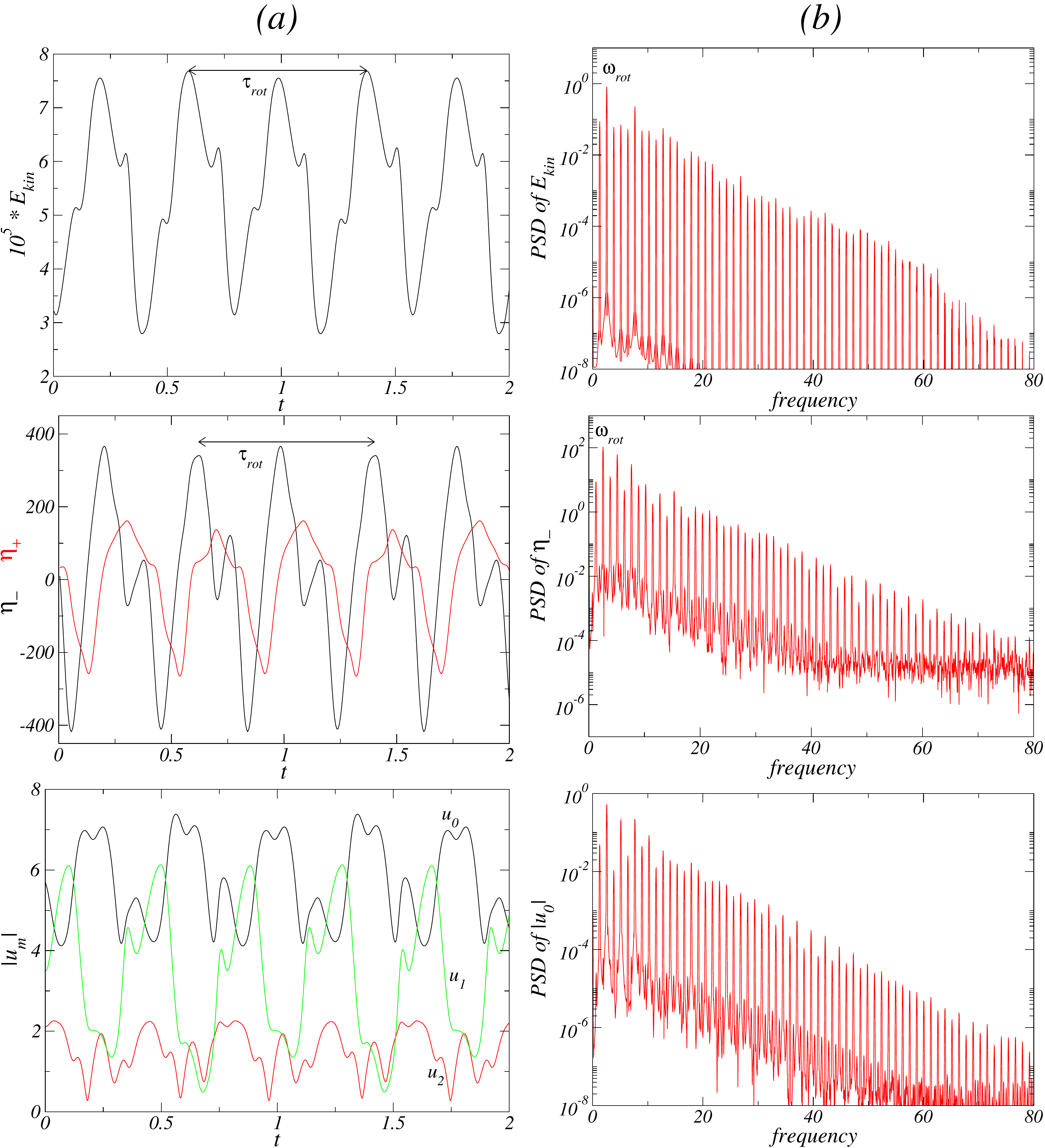}
\end{center}
\caption{{\bf Time series and PSD of the rotating flow state 
$M_{1,2}^\text{rot}$.} For $\Gamma = 1.3$ and $Re_2=-250$, 
$(a)$ Time series of $E_{kin}$, $\eta_+$ [red (gray)], $\eta_-$ (black), and 
the amplitudes $|u_m|$, and $(b)$ the corresponding PSDs. The period of 
rotation is $\tau_\text{rot} \approx 0.7829$ with the corresponding frequency 
$\omega_\text{rot} \approx 2.554$. The frequency of the underlying 
azimuthal oscillation is $\omega_\theta$ (period $\tau_\theta$), which is
visible on the top of the long rotation period with about 
$\tau_\theta \approx \tau_{rot}/8$ (cf. Fig.~\ref{fig:G1_3_R2-250_iso_rv_r-z}).
}
\label{fig:time_PSD_G1_3_M12rot}
\end{figure}

\begin{figure}[]
\begin{center}
\includegraphics[width=1.0\linewidth]{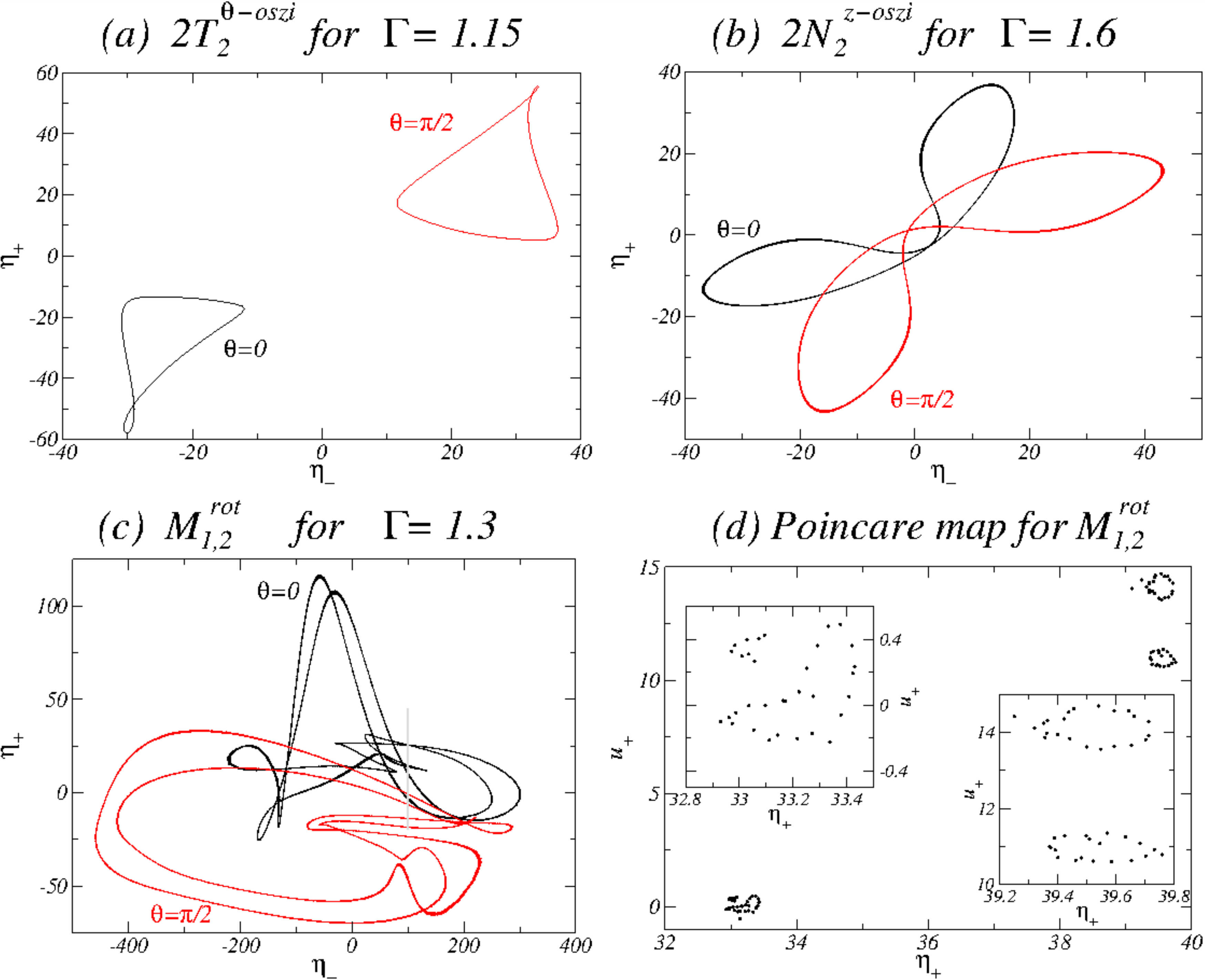}
\end{center}
\caption{{\bf Phase portraits of the flow states for $Re_2=-250$.}
Phase portraits in the $(\eta_+,\eta_-)$ plane of 
$(a)$ $2T_2^{\theta \text{-osci}}$ for $\Gamma=1.15$, 
$(b)$ $2N_2^\text{z-osci}$ for $\Gamma=1.6$, and $(c)$ $M_{1,2}^\text{rot}$ 
for $\Gamma=1.3$, where $\eta_\pm=\eta(r_i,\theta,\pm\Gamma/4,t)$,
and $u_+=u(d/2,0,\Gamma/4,t)$. Black [red (gray)] curves correspond to 
the azimuthal position $\theta=0$ [$\theta=\pi/2$]. $(d)$ The corresponding
two-dimensional Poincar\'{e} section ($u_+,\eta_+$) at $\eta_-=200$
for $\theta=0$ (gray line in $(c)$). The insets provide zoom-in views of 
the shown section to highlight the 2-tori characteristics of the flow 
state $M_{1,2}^\text{rot}$.}
\label{fig:phasespace_R2-250_eta+_vs_eta-}
\end{figure}

At the bifurcation point of $2T_2^{\theta \text{-osci}}$, the kinetic 
energy $\overline{E}_{kin}$ and the axisymmetric mode component $u_0$ 
(cf., Fig.~\ref{fig:bif_scenarios}$(3)$) increase significantly. 
In the mean time the component $u_2$ does not show any significant 
variation in its amplitude. Note that, up to this point, no modes other
than the axisymmetric ($m=0$) and the magnetic field induced ($m=2$) modes
have been stimulated/finite. This picture changes as $\Gamma$ is increased 
further. In particular, for $\Gamma \approx 1.21$, the 
$2T_2^{\theta \text{-osci}}$ mode loses its stability when another 
non-axisymmetric ($m=1$) mode emerges with a second {\em incommensurate} 
frequency $\omega_\text{rot} \approx 2.48$ at the onset.
The flow starts to {\em rotate} 
in the azimuthal direction, so the new state $M_{1,2}^\text{rot}$ is  
a quasiperiodic attractor living on a 2-torus invariant manifold. $M_{1,2}^\text{rot}$
represents a relatively complex flow state in the sense that, within 
one period, the state can exhibit two, three and four cells, as shown in 
Fig.~\ref{fig:G1_3_R2-250_iso_rv_r-z} 
[see also SMs: movieD1.avi and movieD2.avi]. 
Responsible for the rotation of the flow state is the $m=1$ mode 
contribution, which becomes finite at the bifurcation point. This is 
illustrated in Fig.~\ref{fig:time_PSD_G1_3_M12rot}, which includes time 
series and the corresponding PSDs for the stimulated mode amplitudes.
Regarding the time series, PSD in Fig. \ref{fig:time_PSD_G1_3_M12rot} as
phase space plots (cf., Figs.~\ref{fig:phasespace_R2-250_eta+_vs_eta-}$(c,d)$)
the increased complexity of $M_{1,2}^\text{rot}$ to former flow states is obious.

From the $rv$ isosurfaces and the cross-section plots of 
$v(r,\theta,z=1/4[1/2]\Gamma)$ in Fig.~\ref{fig:G1_3_R2-250_iso_rv_r-z},
the strength $u_1$ of the $m=1$ mode contribution can be seen. Comparing 
with the time series and PSDs of different characterizing quantities for the 
$M_{1,2}^\text{rot}$ state (cf., Fig.~\ref{fig:time_PSD_G1_3_M12rot}), 
the $2T^{\theta \text{-osci}}_2$ state 
(cf., Fig.~\ref{fig:time_PSD_G1_15_R2-250_2T2osci}) possesses a higher
degree of complexity. The newly emerged frequency $\omega_\text{rot}$ 
is about 8 times the frequency $\omega_\theta$, which is visible on top 
of the longer rotation period (cf., Fig.~\ref{fig:time_PSD_G1_3_M12rot}).

As $\Gamma$ is increased so that the flow state changes from 
$2T_2^{\theta \text{-osci}}$ to $M^{rot}_{1,2}$, the kinetic energy 
$\overline{E}_{kin}$ continues to increase but it is significantly
smaller than that for the $2T_2^{\theta\text{-osci}}$ state 
(cf., Fig.~\ref{fig:bif_scenarios}$(2)$). In the meantime, the 
emergence (from zero) and enhancement of the $m=1$ mode ($u_1$) is 
compensated by a decrease in the strength of the axisymmetric $m=0$ mode ($u_0$),
with that of the $m=2$ mode ($u_2$) decreasing only slightly. As for 
$Re_2=0$, there exists a regime, $1.18 \lesssim \Gamma \lesssim 1.34$,
in which the non-connected solutions for one- and two-cell flow states
bistable coexist.

Figure~\ref{fig:phasespace_R2-250_eta+_vs_eta-} shows the phase portraits 
of the different time-dependent flow states $2T_2^{\theta \text{-osci}}$, 
$2N_2^\text{z-osci}$ and $M_{1,2}^\text{rot}$ on the $(\eta_+,\eta_-)$ 
plane for $Re_2=-250$ and values of $\Gamma $ as indicated. A Poincar\'{e} 
section $(u_+,\eta_+)$ is also used to better visualize the quasiperiodic 
nature of the flow state $M_{1,2}^\text{rot}$. We see that only the 
oscillating states $2T_2^{\theta \text{-osci}}$ and $2N_2^\text{z-osci}$
possess limit cycle characteristics while the Poincar\'{e} section 
$(u_+,\eta_+)$ highlights the two-tori characteristics of the 
$M_{1,2}^\text{rot}$ state with two incommensurate frequencies, i.e., the 
azimuthal oscillation frequency $\omega_\theta$ and the rotation frequency 
$\omega_\text{rot}$. Note that the curves in $(u_+,\eta_+)$ 
(Fig.~\ref{fig:phasespace_R2-250_eta+_vs_eta-}$(d)$) do not
fully close on themselves due to the long returning times to the section. 
The insets present a zoom-in view of the parameter region. The phase 
portraits illustrate the shift-reflect symmetry of the $2N_2^\text{z-osci}$
state. From the phase portrait for $M_{1,2}^\text{rot}$, an increased 
level of complexity due to the additional rotation of the flow state
can be seen.

\subparagraph*{The case of $Re_2=-500$.}
We now study the case of strongly counter rotating cylinders: $Re_2=-500$. 
The bifurcation diagram with $\Gamma$ is shown in 
Fig.~\ref{fig:bif_scenarios}$(4)$. For $\Gamma=1.6$ we find the flow state $2N_2$
(Fig.~\ref{fig:bif_scenarios}$(1)$) 
Increasing $\Gamma$, the state loses its stability at $\Gamma\approx 1.62$ 
through a symmetry breaking Hopf bifurcation and the resulting state is 
again the unsteady, axially oscillating flow state $2N^\text{z-osci}_2$
(cf., Figs.~\ref{fig:G1_6_R2-250_iso_rv_r-z} and 
\ref{fig:time_PSD_G1_6_R2-250_2N2osci}), which remains stable until 
$\Gamma \approx 1.75$.

We find that the oscillation 
amplitude is much smaller than that for the $2N_2^\text{z-osci}$ state at 
$Re_2=-250$ (cf. Fig.~\ref{fig:G1_6_R2-250_iso_rv_r-z}).
As $\Gamma$ is decreased, the state remains stable until for $\Gamma\approx 1.18$
when it loses stability, at which the one-cell flow state $1A_2$ 
($1A_2^*$) emerges, similar to the cases of $Re_2=0$ and $Re_2=-250$.
In the opposite direction, i.e., starting from the steady flow state 
$2N_2^\text{c}$ for $\Gamma=0.5$ and increasing $\Gamma$, the state 
loses its stability at $\Gamma \approx 0.59$ through the same symmetry 
breaking pitchfork bifurcation as for the cases of $Re_2=0$ and $Re_2=-250$. 
Due to the stronger counter rotation, two additional vortex cells start 
to develop near the outer cylinders for the flow state $2N_2^\text{c}$.
The symmetry related one-cell flow state $1A_2$ or $1A_2^*$ appears
as the size of one of the vortex cells increases or shrinks, respectively. 
Upon further increase in $\Gamma$, the flow state $1A_2$ remains stable 
with slight but continuous change in the position of the vortex cell. The 
larger vortex cell moves towards the inner cylinder, while the second 
vortex cell grows and moves radially outward towards the outer cylinder.
In principle this is the same evolution as for the case of $Re_2=-250$, 
with the only difference being that, due to the stronger counter rotation
($Re_2=-500$), the vortex cells and in particular their centers are 
slightly shifted and located closer towards the inner cylinder.

For $\Gamma \approx 1.08$, the flow becomes time dependent, and the 
azimuthally oscillating twin-cell flow state $2T^{\theta \text{-osci}}_2$
emerges.
The dynamics is almost the same as seen in
Figs~\ref{fig:G1_15_R2-250_iso_rv_r-z} and 
\ref{fig:time_PSD_G1_15_R2-250_2T2osci} for slightly smaller $\Gamma=1.15$ and
$Re_2=-250$. With a further increase in $\Gamma$, this state loses its stability at 
$\Gamma \approx 1.35$ and becomes a transient to the flow state $2N_2$. 
For $Re_2=-500$, there then exists again a regime, $1.18\lesssim\Gamma\lesssim1.34$,
in which the one-cell and two-cell flow states bistable coexist. Differing from 
the case of $Re_1=-250$, there is absence of more complex (e.g., 
quasiperiodic solution) flow state for $Re_2=-500$.
The phase portraits for the azimuthally and axially oscillating flow 
states $2N_2^\text{z-osci}$ and $2T_2^{\theta\text{-osci}}$ is similar
to these solutions at $Re_2=-250$.

A focus of our present study is axially or azimuthally oscillating flow 
states. It should be noted, however, that flow states of {\em combined} 
axial and azimuthal oscillations can occur in the parameter regime of 
larger aspect ratio and very large values of the Reynolds number.
A more detailed discussion about the behaviors of the angular momentum and 
torque can be found in Supplementary Materials. \\
\\ \noindent 
{\large\bf Discussion and summary of main findings} \\
\\ \noindent
As a fundamental paradigm of fluid dynamics, the TCS has been 
extensively investigated computationally and experimentally. In spite
of the long history of the TCS and the vast literature on the subject, 
the dynamics of TCS with a complex fluid subject to a symmetry breaking 
magnetic field have begun to be investigated relatively recently. 
In fact, a gap existed in our knowledge about the nonlinear dynamics
of such systems with a small aspect ratio. The present work is
aimed to fill this gap. 

Through systematic and extensive simulations 
of the ferrohydrodynamical equations, a generalization of the classic
Navier-Stokes equation into ferrofluidic systems subject to a magnetic 
field, we unveil the emergence and evolution of distinct dynamical
flow states. As the Reynolds number of the outer cylinder and/or the 
aspect ratio is changed, symmetry-breaking pitchfork and Hopf 
bifurcations can occur, leading to transitions among various flow
states, e.g., the two-cell and one-cell states. The presence of the 
transverse magnetic field stipulates that all flow states must inherently 
be three-dimensional~\cite{SACHALML2010,ReOd:2011,ALD:2013}. 

The detailed emergence, dynamical evolution, and transitions among the
various flow states can be summarized, as follows. We first identify 
a fundamental building block that plays a dominant role in the formation
of various flow structures: the order-two azimuthal $m=2$ mode. For 
small aspect ratio (e.g., $\Gamma \approx 0.5$), the two-cell state 
$2N_2^\text{c}$ dominates which, due to its two-fold flow symmetry, 
differs little from the one in the classical TCS~\cite{BSP:1992}. 
Depending on the rotational speed and the direction of the outer 
cylinder the vortex cells within the $2N_2^\text{c}$ state can move 
closer towards the inner cylinder. The flow is steady 
and exhibits a more complex set of symmetries associated with the 
magnetic field, namely, a combination of the two-fold symmetry and 
the reflection symmetry about the mid-plane under reversal of the field 
direction. As $\Gamma$ is increased, this flow state undergoes a 
symmetry breaking bifurcation at which one vortex cell starts to grow
while the other begins to shrink, eventually generating two symmetry 
related one-cell flow states: $1A_2$ and $1A_2^*=K^H_z1A_2$. When the 
outer cylinder is at rest (i.e., $Re_2 = 0$), the state $1A_2$ 
loses stability and eventually becomes a transient to the steady, axially 
symmetric two-cell flow state $2N_2$. For counter-rotating cylinders 
(i.e., $Re_2 < 0$), we find a transition to the same flow state $2N_2$ 
at a larger value of $\Gamma$. However, prior to this transition a 
distinct bifurcation sequence leading to {\em new unsteady} flow states
occurs. In fact, as $\Gamma$ is increased, the one-cell flow state $1A_2$ 
becomes modulated in that the smaller vortex cell grows along the inner 
cylinder while the other vortex cell is pulled outward. Eventually the 
steady flow state $1A_2$ undergoes a Hopf bifurcation to a periodic, 
{\em azimuthally oscillating} flow state $2T_2^{\theta \text{-osci}}$ in 
the twin-cell configuration (side-by-side arrangement)
where both vortex cells touch the top and 
bottom lids, which topologically corresponds to a limit cycle.
During the dynamical evolution, there are two symmetry related flow 
states: $2T_2^{\theta \text{-osci}}$ and 
$2T_2^{\theta \text{-osci},*}=K^H_z2T_2^{\theta \text{-osci}}$.
Increasing $\Gamma$ further, we find two possible bifurcation scenarios.
First, $2T_2^{\theta \text{-osci}}$ loses its stability and becomes a 
transient to the steady flow state $2N_2$. Second, 
$2T_2^{\theta \text{-osci}}$ becomes unstable, leading to an unsteady 
quasiperiodic flow state $M_{1,2}^\text{rot}$. The quasiperiodic state 
has finite contribution from the $m=1$ mode and rotates in the azimuthal 
direction. As a result, one of the two frequencies, $\omega_\theta$, 
corresponds to the frequency of the underlying $2T_2^{\theta \text{-osci}}$
mode from which it bifurcates, and the second frequency $\omega_\text{rot}$
comes from the rotation of the $m=1$ mode, a flow state with
a helical shape (cf., Fig.~\ref{fig:G1_3_R2-250_iso_rv_r-z}). For larger 
values of $\Gamma$, the unsteady flow state $M_{1,2}^\text{rot}$ eventually
loses its stability and becomes transient towards the steady state $2N_2$.

A similar scenario occurs when the aspect ratio is varied in the opposite
direction, i.e., from large to small values. At a certain point the steady 
two-cell flow state $2N_2$ loses its stability and is replaced by one of 
the symmetry related steady one-cell flow states, $1A_2$ or $1A_2^*$. 
Depending on the parameters there is a relatively large regime in which the both
not-connected solutions, two-cell and one-cell flow states bistable coexist.
It is worth mentioning that 
our computations never reveal any signature of the transition from the 
steady two-cell flow state $2N_2$ to any of the unsteady one-cell flow 
states (i.e., $2T_2^{\theta \text{-osci}}$ or $M_{1,2}^\text{rot}$). 
The reduction in the vortex cells (from two to one) appears to happen
only between the steady flow states. 

In addition to these unsteady flow states, we detect another unsteady 
flow state, the axially oscillating flow state $2N_2^\text{z-osci}$ that
is known for the classical TCS~\cite{BSP:1992}.
The state $2N_2^\text{z-osci}$ emerges at a large value of $\Gamma$ or 
through variation of the rotation speed of the outer cylinder through a 
symmetry breaking Hopf bifurcation out of the flow state $2N_2$. Similar
to $2T_2^{\theta \text{-osci}}$, the flow state $2N_2^\text{z-osci}$
is a limit cycle solution which is half-period flip invariant under the 
symmetry operation $S^H$.
 
\begin{figure}[]
\begin{center}
\includegraphics[width=1.0\linewidth]{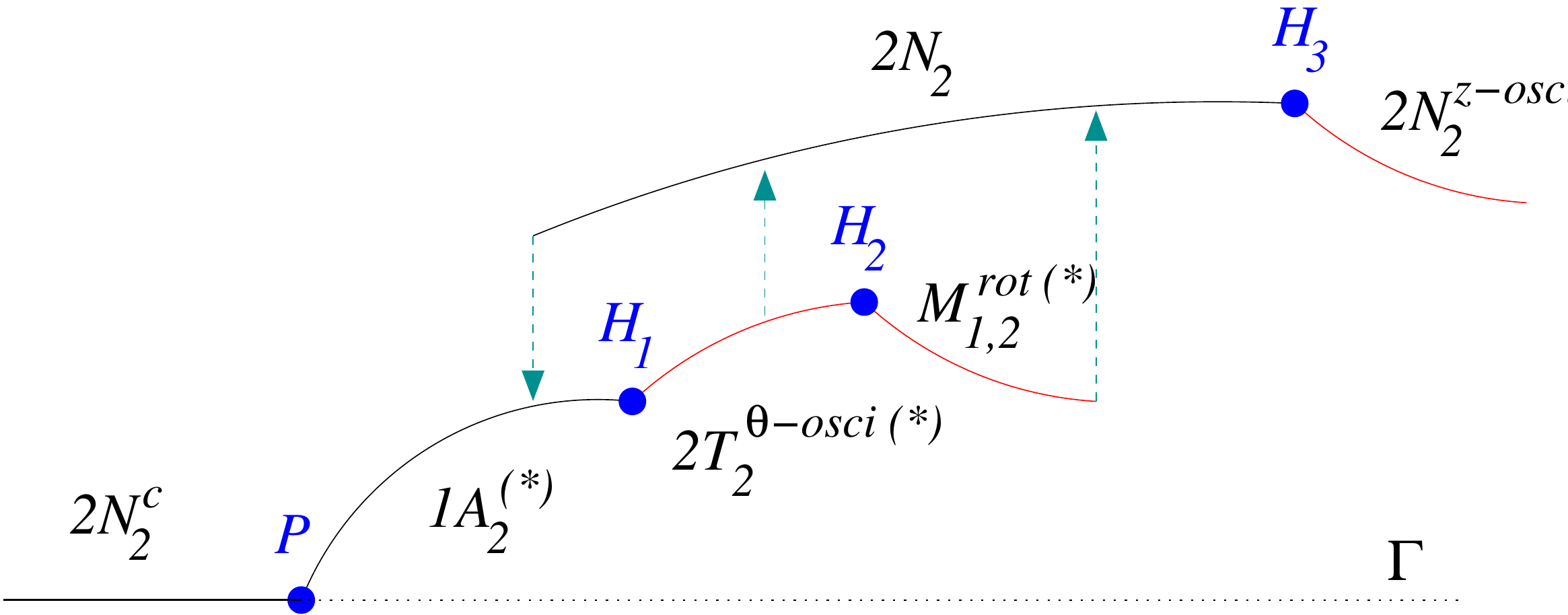}
\end{center}
\caption{{\bf Summary: schematic bifurcations for ferrofluidic TCS with a 
small aspect ratio.} For illustrative purpose, the aspect ratio $\Gamma$ 
is taken as the bifurcation parameter, where $P$ denotes the pitchfork 
bifurcation of $2N_2^\text{c}$ into two symmetry related flow states $1A_2$
and $1A_2^*=K_z^H 1A_2$. The Hopf bifurcation $H_1$ generates the limit 
cycle solution $2T_2^{\theta \text{-osci}}$ (or 
$2T_2^{\theta \text{-osci},*}$). At the Hopf bifurcation $H_2$, a 
quasiperiodic symmetric flow state $M_{1,2}^\text{rot}$ (or 
$M_{1,2}^{\text{rot},*}$) is born out of the limit cycle. Depending on 
other system parameters the flow becomes transient towards the $2N_2$ 
state from either $2T_2^{\theta \text{-osci}}$ or $M_{1,2}^\text{rot}$. 
The state $2N_2$ undergoes another Hopf bifurcation $H_3$, generating 
a distinct limit cycle solution, $2N_2^\text{z-osci}$. Black [Red (gray)] 
colored lines indicate steady [unsteady] flow states.}
\label{fig:schematics}
\end{figure}

To summarize the complicated bifurcation/transition scenarios in 
the ferrofluidic TCS with a small aspect ratio in a transparent way, we 
produce a schematic bifurcation diagram with the aspect ratio $\Gamma$
being the bifurcation parameter, as shown in Fig.~\ref{fig:schematics}.
The symmetry breaking associated with each bifurcation point can be 
described succinctly, as follows. At the pitchfork bifurcation point $P$,
the two-cell flow state $2N_2^\text{c}$, which is invariant under the 
symmetries $R_\pi^H$ and $K_z^H$, loses stability, giving rise to the 
one-cell flow state $1A_2$. In fact, breaking the $K^H_z$ symmetry
results in two symmetry related flow states $1A_2$ and $1A_2^*=K^H_z1A_2$. 
The state $1A_2$ loses stability through the Hopf bifurcation $H_1$ at
which a limit cycle state $2T_2^{\theta \text{-osci}}$ (or 
$2T_2^{\theta \text{-osci},*}$) is born. Finally a second frequency 
appears through another Hopf bifurcation $H_2$, leading to a two-frequency 
quasiperiodic solution $M_{1,2}^\text{rot}$ (or $M_{1,2}^{\text{rot},*}$).\\
\\ \noindent
{\large\bf Methods}
\\
\paragraph*{System setting and the Navier-Stokes equation.}
We consider a standard TCS consisting of two concentric, independently 
rotating cylinders. Within the gap between the two cylinders there is an 
incompressible, isothermal, homogeneous, mono-dispersed ferrofluid of 
kinematic viscosity $\nu$ and density $\rho$. The inner and outer 
cylinders have radius $R_1$ and $R_2$, and they rotate with the angular 
velocity $\omega_1$ and $\omega_2$, respectively. The boundary conditions at 
the cylinder surfaces are of the non-slip type, and the end walls enclosing 
the annulus are stationary. The system can be characterized in the 
cylindrical coordinate system $(r,\theta,z)$ by the velocity field 
${\bf u}=(u,v,w)$ and the corresponding vorticity field 
$\nabla \times \bf{u} = (\xi,\eta,\zeta)$. We fix the radius ratio of the 
cylinders: $R_1/R_2=0.5$, and vary the height-to-gap aspect ratio of the 
annulus in the range $0.5 \leqslant \Gamma \leqslant 1.75$. A homogeneous 
magnetic field ${\bf H} = H_x{\bf e}_x$ is applied in the transverse 
($x=r\cos{\theta}$) direction, with $H_x$ being the field strength. The
length and time scales of the system are set by the gap width $d = R_2 - R_1$ 
and the diffusion time $d^2/\nu$, respectively. The pressure in the fluid
is normalized by $\rho\nu^2/d^2$, and the magnetic field ${\bf H}$ and the
magnetization ${\bf M}$ can be conveniently normalized by the quantity 
$\sqrt{\rho/\mu_0} \nu/d$, where $\mu_0$ is the permeability of free space. 
These considerations lead to the following set of non-dimensionalized 
hydrodynamical equations~\cite{ALD:2013,MuLi:2001}:
\begin{eqnarray}
\nonumber
(\partial_t + {\bf u \cdot \nabla}) {\bf u} - \nabla^2 {\bf u}+ \nabla p 
&=& ({\bf M}\cdot \nabla) {\bf H} + 
\frac{1}{2} \nabla \times ({\bf M}\times {\bf H}), \\
\label{EQ:nast}
\nabla \cdot {\bf u} &=& 0.
\end{eqnarray}
The boundary conditions are set as follows. The velocities at the stationary 
boundaries (i.e., lids) are zero. On the cylindrical surfaces, the velocity
fields are given by ${\bf u}(r_1,\theta,z)=(0,Re_1,0)$ and 
${\bf u}(r_2,\theta,z)=(0,Re_2,0)$, where the inner and outer Reynolds 
numbers are $Re_1=\omega_1 r_1 d/\nu$ (fixed at 250 in the present study)
and $Re_2=\omega_2 r_2 d/\nu$, respectively, where $r_1=R_1/(R_2-R_1)$ and 
$r_2=R_2/(R_2-R_1)$ are the non-dimensionalized inner and outer cylinder 
radii, respectively.
Note that the idealized boundary conditions are discontinuous at the 
junctions where the stationary end walls meet the rotating cylinders. In 
experiments there are small but finite gaps at these junctions where the 
azimuthal velocity adjusts to zero. To treat the boundary conditions  
properly, we implement the following regularization procedure for the 
boundary conditions:
\begin{eqnarray}
v(r,\theta\pm0.5\Gamma,t) = Re_1 exp([r_1 - r]/\epsilon) + 
Re_2 exp([r - r_2]/\epsilon),
\end{eqnarray}
where $\epsilon$ is a small parameter characterizing the physical
gaps. In the simulations, we set $\epsilon = 6 \times 10^{-3}$.
The range of variation in $Re_2$ is  $-2000 \le Re_2 \le 500$.

\paragraph*{Ferrohydrodynamical equation.}
Equation~\eqref{EQ:nast} is to be solved together with an equation that
describes the magnetization of the ferrofluid. Using the equilibrium 
magnetization of an unperturbed state in which the homogeneously magnetized 
ferrofluid is at rest and the mean magnetic moment is orientated in the 
direction of the magnetic field, we have ${\bf M^\text{eq}} = \chi {\bf H}$. 
The magnetic susceptibility $\chi$ of the ferrofluid can be approximated 
by the Langevin's formula~\cite{Lan:1905}, where we set the initial value of 
$\chi$ to be 0.9 and use a linear magnetization law. The ferrofluid studied 
corresponds to APG933~\cite{EMWKL:2000}. We consider the near equilibrium 
approximations of Niklas~\cite{Nik:1987,NML:1989} with a small value of 
$||{\bf M} - {\bf M}^\text{eq}||$ and small magnetic relaxation time $\tau$: 
$|\nabla \times {\bf u}| \tau \ll 1$. Using these approximations, one can 
obtain~\cite{ALD:2013} the following magnetization equation:
\begin{equation} \label{EQ:niklas}
{\bf M} - {\bf M}^\text{eq} = c^2_N 
\left( \frac{1}{2} \nabla \times {\bf u} \times {\bf H} 
+ \lambda_2  {\mathbb S} {\bf H} \right),   
\end{equation}
where
\begin{equation}
c^2_N =\tau / \left(\displaystyle 1/\chi + \displaystyle \tau
\mu_0 H^2 / 6\mu\Phi\right)
\end{equation}
is the Niklas coefficient~\cite{Nik:1987}, $\mu$ is the dynamic viscosity, 
$\Phi$ is the volume fraction of the magnetic material, ${\mathbb S}$ is the 
symmetric component of the velocity gradient tensor~\cite{MuLi:2001,ALD:2013}, 
and $\lambda_2$ is the material-dependent transport 
coefficient~\cite{MuLi:2001} that can be conveniently chosen to 
be~\cite{AHLL:2010,MuLi:2001,OdMu:2002} $\lambda_2 = 4/5$.
Using Eq.~\eqref{EQ:niklas}, we eliminate the magnetization from 
Eq.~\eqref{EQ:nast} to arrive at the following ferrohydrodynamical 
equations~\cite{MuLi:2001,ALD:2013}:
\begin{eqnarray} \label{FHD}  
(\partial_{t} + {\bf u}\cdot \nabla ) {\bf u} - \nabla^{2}{\bf u} 
+ \nabla p_M \qquad \\ \nonumber
=  - \frac{ s^2_x}{2} \left[{\bf H} \nabla \cdot 
\left({\bf F} + \frac{4}{5} {\mathbb S} {\bf H}\right) + {\bf H} \times 
\nabla \times \left({\bf F} + \frac{4}{5} {\mathbb S} {\bf H}\right)\right],
\end{eqnarray}
where ${\bf F} = (\nabla \times {\bf u}/2) \times {\bf H}$, $p_M$ is the 
dynamic pressure incorporating all magnetic terms that can be expressed as 
gradients, and $s_x$ is the Niklas parameter [Eq.~\eqref{EQ:niklas_param}]. 
To the leading order, the internal magnetic field in the ferrofluid can be 
approximated by the externally imposed field~\cite{ALD:2012}, which is 
reasonable for obtaining the dynamical solutions of the magnetically driven 
fluid motion. Equation~\eqref{FHD} can then be simplified as
\begin{eqnarray} \label{EQ:NSE_without_M_special}
(\partial_t + {\bf u \cdot \nabla}) {\bf u} - \nabla^{2} {\bf u} + 
\nabla p_M  =  s_x^2 \left\lbrace \nabla^2 {\bf u} - \frac{4}{5} 
\left[ \nabla \cdot ({\mathbb S} {\bf H}) \right] \right. \qquad \\ \nonumber 
- {\bf H} \times \left[\frac{1}{2} \nabla \times (\nabla \times {\bf u} 
\times {\bf H}) - {\bf H} \times (\nabla^2 {\bf u}) \right. 
\left. \left. + \frac{4}{5} \nabla \times ({\mathbb S} {\bf H}) 
\right] \right\rbrace .
\end{eqnarray}
This way, the effect of the magnetic field and the magnetic properties of 
the ferrofluid on the velocity field can be characterized by a single 
parameter, the magnetic field or the Niklas parameter~\cite{Nik:1987}:
\begin{equation} \label{EQ:niklas_param}
s_x^2 = \cfrac{2(2+\chi)H_x c_N}{(2+\chi)^2-\chi^2\eta^2}.
\end{equation}

\paragraph*{Numerical method.}
The ferrohydrodynamical equations of motion Eq.~\eqref{FHD} can be 
solved~\cite{AHLL:2010,ALD:2012,ALD:2013} by combining a standard, 
second-order finite-difference scheme in $(r,z)$ with a Fourier spectral 
decomposition in $\theta$ and (explicit) time splitting. 
The variables can be expressed as
\begin{equation}
f(r,\theta,z,t)=\sum_{m=-m_{\max}}^{m_{\max}} f_m(r,z,t)\,e^{im\theta},
\end{equation}
where $f$ denotes one of the variables $\{u,v,w,p\}$. For the parameter 
regimes considered, the choice $m_{\max}=10$ provides adequate accuracy. 
We use a uniform grid with spacing $\delta r = \delta z =0.02$ and time 
steps $\delta t < 1/3800$. 

\paragraph*{Symmetries.}
In a classical TCS or a ferrofluidic TCS without any external magnetic field 
where the fluid is confined by end walls, the system is invariant with respect 
to arbitrary rotations about the axis and the reflections about axial 
mid-height. For a ferrofluid under a transverse magnetic field, 
these symmetries are broken and the flow is inherently three-dimensional 
for any non-zero values of the parameters $Re_1$, $Re_2$ and $s_x$, due to 
the rotation of the cylinders~\cite{SACHALML2010,ReOd:2011,ALD:2012,ALD:2013}.
With at least one cylinder rotating, the inclusion of the magnetic
terms in the ferrohydrodynamic equation results in a downward directed
force on the side where the field enters the system ($\theta=0$), and an 
upward directed force on the opposite side ($\theta=\pi$) where the field 
exits the annulus. The resulting flow states can possess more complicated
symmetries, such as the reflection $K^H_z$ about the annulus mid-height 
plane along with an inversion of the magnetic field direction. There can
also be a rotational invariance $R^H_\alpha$ for discrete angle $\alpha=\pi$ 
in combination with the reversal of the magnetic field, where the angle 
$\pi$ specifies the direction of the magnetic field when entering the 
annulus. Thus the symmetries associated with the velocity field are 
\begin{eqnarray}
\nonumber
R^H_{\pi}(u,v,w,H)(r,\theta,z,t)&=&(u,v,w,-H)(r,\theta+\pi,z,t), \\
\label{EQ:symmetry} \\
\nonumber
K^H_z(u,v,w,H)(r,\theta,z,t)&=&(u,v,-w,-H)(r,\theta,-z,t).
\end{eqnarray}
For a periodic solution (with period $\tau$), the flow field is also
invariant under the discrete time translation
\begin{eqnarray}
\nonumber
\Phi_{\tau}(u,v,w,H)(r,\theta,z,t)&=&(u,v,w,H)(r,\theta,z,t+\tau).
\end{eqnarray}
Further details of the magnetic field induced two-fold symmetry can be found
in Ref.~\cite{ALD:2013}.


\begin{thebibliography}{10}
\expandafter\ifx\csname url\endcsname\relax
  \def\url#1{\texttt{#1}}\fi
\expandafter\ifx\csname urlprefix\endcsname\relax\def\urlprefix{URL }\fi
\providecommand{\bibinfo}[2]{#2}
\providecommand{\eprint}[2][]{\url{#2}}

\bibitem{Tay:1923}
\bibinfo{author}{Taylor, G.~I.}
\newblock \bibinfo{title}{Stability of a viscous liquid contained between two
  rotating cylinders}.
\newblock \emph{\bibinfo{journal}{Philos. Trans. R. Soc. London A}}
  \textbf{\bibinfo{volume}{223}}, \bibinfo{pages}{289} (\bibinfo{year}{1923}).

\bibitem{ChIo:1994}
\bibinfo{author}{Chossat, P.} \& \bibinfo{author}{Iooss, G.}
\newblock \emph{\bibinfo{title}{The Couette--Taylor Problem}}
  (\bibinfo{publisher}{Springer}, \bibinfo{address}{Berlin},
  \bibinfo{year}{1994}).

\bibitem{AHLL:2010}
\bibinfo{author}{Altmeyer, S.}, \bibinfo{author}{Hoffmann, C.},
  \bibinfo{author}{Leschhorn, A.} \& \bibinfo{author}{L\"ucke, M.}
\newblock \bibinfo{title}{Influence of homogeneous magnetic fields on the flow
  of a ferrofluid in the taylor-couette system}.
\newblock \emph{\bibinfo{journal}{Phys.~Rev.~E}} \textbf{\bibinfo{volume}{82}},
  \bibinfo{pages}{016321} (\bibinfo{year}{2010}).

\bibitem{ReOd:2010}
\bibinfo{author}{Reindl, M.} \& \bibinfo{author}{Odenbach, S.}
\newblock \bibinfo{title}{Influence of a homogeneous axial magnetic field on
  {T}aylor-{C}ouette flow of ferrofluids with low particle-particle
  interaction}.
\newblock \emph{\bibinfo{journal}{Expts.~Fluids}}
  \textbf{\bibinfo{volume}{50}}, \bibinfo{pages}{375--384}
  (\bibinfo{year}{2011}).

\bibitem{ReOd:2011}
\bibinfo{author}{Reindl, M.} \& \bibinfo{author}{Odenbach, S.}
\newblock \bibinfo{title}{Effect of axial and transverse magnetic fields on the
  flow behavior of ferrofluids featuring different levels of interparticle
  interaction}.
\newblock \emph{\bibinfo{journal}{Phys.~Fluids}} \textbf{\bibinfo{volume}{23}},
  \bibinfo{pages}{093102} (\bibinfo{year}{2011}).

\bibitem{Alt2011}
\bibinfo{author}{Altmeyer, S.}
\newblock \emph{\bibinfo{title}{Untersuchungen von komplexen Wirbelstr\"omungen
  mit newtonschem Fluid und Ferrofluiden im Taylor-Couette System}}.
\newblock \bibinfo{type}{Doktorarbeit}, \bibinfo{school}{Universit\"at des
  Saarlandes}, \bibinfo{address}{Saarbr\"ucken} (\bibinfo{year}{2011}).
\newblock \bibinfo{note}{Unver\"offentlicht}.

\bibitem{ALD:2013}
\bibinfo{author}{Altmeyer, S.}, \bibinfo{author}{Lopez, J.} \&
  \bibinfo{author}{Do, Y.}
\newblock \bibinfo{title}{Effect of elongational flow on ferrofuids under a
  magnetic field}.
\newblock \emph{\bibinfo{journal}{Phys.~Rev.~E}} \textbf{\bibinfo{volume}{88}},
  \bibinfo{pages}{013003} (\bibinfo{year}{2013}).

\bibitem{ADL:2015a}
\bibinfo{author}{Altmeyer, S.}, \bibinfo{author}{Do, Y.-H.} \&
  \bibinfo{author}{Lai, Y.-C.}
\newblock \bibinfo{title}{Transition to turbulence in taylor-couette
  ferrofluidic flow}.
\newblock \emph{\bibinfo{journal}{Sci. Rep.}} \textbf{\bibinfo{volume}{5}},
  \bibinfo{pages}{10781} (\bibinfo{year}{2015}).

\bibitem{ADL:2015b}
\bibinfo{author}{Altmeyer, S.}, \bibinfo{author}{Do, Y.-H.} \&
  \bibinfo{author}{Lai, Y.-C.}
\newblock \bibinfo{title}{Magnetic field induced flow reversal in a
  ferrofluidic taylor-couette system}.
\newblock \emph{\bibinfo{journal}{Sci. Rep.}} \textbf{\bibinfo{volume}{5}},
  \bibinfo{pages}{18589} (\bibinfo{year}{2015}).

\bibitem{Ros:1985}
\bibinfo{author}{Rosensweig, R.~E.}
\newblock \emph{\bibinfo{title}{Ferrohydrodynamics}}
  (\bibinfo{publisher}{Cambridge University Press},
  \bibinfo{address}{Cambridge}, \bibinfo{year}{1985}).

\bibitem{Mc69}
\bibinfo{author}{McTague, J.~P.}
\newblock \bibinfo{title}{Magnetoviscosity of magnetic colloids}.
\newblock \emph{\bibinfo{journal}{J.~Chem.~Phys.}}
  \textbf{\bibinfo{volume}{51}}, \bibinfo{pages}{133} (\bibinfo{year}{1969}).

\bibitem{Sh72}
\bibinfo{author}{Shliomis, M.~I.}
\newblock \bibinfo{title}{Effective viscosity of magnetic suspensions}.
\newblock \emph{\bibinfo{journal}{Sov. Phys. JETP}}
  \textbf{\bibinfo{volume}{34}}, \bibinfo{pages}{1291} (\bibinfo{year}{1972}).

\bibitem{Har06}
\bibinfo{author}{Hart, J.~E.}
\newblock \bibinfo{title}{A magnetic fluid laboratory model of the global
  buoyancy and wind-driven ocean circulation: {A}nalysis}.
\newblock \emph{\bibinfo{journal}{Dyn.~Atmos.~Oceans}}
  \textbf{\bibinfo{volume}{41}}, \bibinfo{pages}{121--138}
  (\bibinfo{year}{2006}).

\bibitem{HaKi:2006}
\bibinfo{author}{Hart, J.~E.} \& \bibinfo{author}{Kittelman, S.}
\newblock \bibinfo{title}{A magnetic fluid laboratory model of the global
  buoyancy and wind-driven ocean circulation: {E}xperiments}.
\newblock \emph{\bibinfo{journal}{Dyn.~Atmos.~Oceans}}
  \textbf{\bibinfo{volume}{41}}, \bibinfo{pages}{139--147}
  (\bibinfo{year}{2006}).

\bibitem{GR:1995a}
\bibinfo{author}{Glatzmaier, G.~A.} \& \bibinfo{author}{Roberts, P.~H.}
\newblock \bibinfo{title}{A three dimensional self-consistent computer
  simulation of a geomagnetic field reversal}.
\newblock \emph{\bibinfo{journal}{Nature}} \textbf{\bibinfo{volume}{377}},
  \bibinfo{pages}{203--209} (\bibinfo{year}{1995}).

\bibitem{GR:1995b}
\bibinfo{author}{Glatzmaier, G.~A.} \& \bibinfo{author}{Roberts, P.~H.}
\newblock \bibinfo{title}{A three-dimensional convective dynamo solution with
  rotating and finitely conducting inner core and mantle}.
\newblock \emph{\bibinfo{journal}{Phys. Earth Planet. Inter.}}
  \textbf{\bibinfo{volume}{91}}, \bibinfo{pages}{63--75}
  (\bibinfo{year}{1995}).

\bibitem{GR:1996}
\bibinfo{author}{Glatzmaier, G.~A.} \& \bibinfo{author}{Roberts, P.~H.}
\newblock \bibinfo{title}{Rotation and magnetism of earth's inner core}.
\newblock \emph{\bibinfo{journal}{Science}} \textbf{\bibinfo{volume}{274}},
  \bibinfo{pages}{1887--1891} (\bibinfo{year}{1996}).

\bibitem{GCHR:1999}
\bibinfo{author}{Glatzmaier, G.~A.}, \bibinfo{author}{Coe, R.~S.},
  \bibinfo{author}{Hongre, L.} \& \bibinfo{author}{Roberts, P.~H.}
\newblock \bibinfo{title}{The role of the earth's mantle in controlling the
  frequency of geomagnetic reversals}.
\newblock \emph{\bibinfo{journal}{Nature}} \textbf{\bibinfo{volume}{401}},
  \bibinfo{pages}{885--890} (\bibinfo{year}{1999}).

\bibitem{GR:2000}
\bibinfo{author}{Glatzmaier, G.~A.} \& \bibinfo{author}{Roberts, P.~H.}
\newblock \bibinfo{title}{Geodynamo theory and simulations}.
\newblock \emph{\bibinfo{journal}{Rev. Mod. Phys.}}
  \textbf{\bibinfo{volume}{72}}, \bibinfo{pages}{1081--1123}
  (\bibinfo{year}{2000}).

\bibitem{Har:2002}
\bibinfo{author}{Hart, J.~E.}
\newblock \bibinfo{title}{Ferromagnetic rotating {C}ouette flow: {T}he role of
  magnetic viscosity}.
\newblock \emph{\bibinfo{journal}{J.~Fluid~Mech.}}
  \textbf{\bibinfo{volume}{453}}, \bibinfo{pages}{21--38}
  (\bibinfo{year}{2002}).

\bibitem{Berhanuetal:2007}
\bibinfo{author}{Berhanu1, M.} \emph{et~al.}
\newblock \bibinfo{title}{Magnetic field reversals in an experimental turbulent
  dynamo}.
\newblock \emph{\bibinfo{journal}{Europhys. Lett.}}
  \textbf{\bibinfo{volume}{77}}, \bibinfo{pages}{59001} (\bibinfo{year}{2007}).

\bibitem{Ben78a}
\bibinfo{author}{Benjamin, T.}
\newblock \bibinfo{title}{Bifurcation phenomena in steady flows of a viscous
  fluid. i. theory.}
\newblock \emph{\bibinfo{journal}{Philo. Trans. Roy. Soc. A}}
  \bibinfo{pages}{1--26} (\bibinfo{year}{1978}).

\bibitem{Ben78b}
\bibinfo{author}{Benjamin, T.}
\newblock \bibinfo{title}{Bifurcation phenomena in steady flows of a viscous
  fluid. ii. experiments.}
\newblock \emph{\bibinfo{journal}{Philo. Trans. Roy. Soc. A}}
  \bibinfo{pages}{27--43} (\bibinfo{year}{1978}).

\bibitem{CKM92}
\bibinfo{author}{Cliffe, K.~A.}, \bibinfo{author}{Kobine, J.~J.} \&
  \bibinfo{author}{Mullin, T.}
\newblock \bibinfo{title}{The role of anomalous modes in {T}aylor-{C}ouette
  flow}.
\newblock \emph{\bibinfo{journal}{Philo. Trans. Roy. Soc. A}}
  \textbf{\bibinfo{volume}{439}}, \bibinfo{pages}{341--357}
  (\bibinfo{year}{1992}).

\bibitem{AHHAPLP2010}
\bibinfo{author}{Altmeyer, S.} \emph{et~al.}
\newblock \bibinfo{title}{End wall effects on the transitions between taylor
  vortices and spiral vortices}.
\newblock \emph{\bibinfo{journal}{Phys.~Rev.~E}} \textbf{\bibinfo{volume}{81}},
  \bibinfo{pages}{066313} (\bibinfo{year}{2010}).

\bibitem{BeMu:1981}
\bibinfo{author}{Benjamin, B.} \& \bibinfo{author}{Mullin, T.}
\newblock \bibinfo{title}{Anomalous modes in the taylor experiment}.
\newblock \emph{\bibinfo{journal}{Proc. R. Soc. London Ser. A}}
  \textbf{\bibinfo{volume}{377}}, \bibinfo{pages}{221--249}
  (\bibinfo{year}{1981}).

\bibitem{LMWP:1984}
\bibinfo{author}{L\"ucke, M.}, \bibinfo{author}{Mihelcic, M.},
  \bibinfo{author}{Wingerath, K.} \& \bibinfo{author}{Pfister, G.}
\newblock \bibinfo{title}{Flow in a small annulus between concentric
  cylinders}.
\newblock \emph{\bibinfo{journal}{J. Fluid Mech.}}
  \textbf{\bibinfo{volume}{140}}, \bibinfo{pages}{343--353}
  (\bibinfo{year}{1984}).

\bibitem{PSCM:1988}
\bibinfo{author}{Pfister, G.}, \bibinfo{author}{Schmidt, H.},
  \bibinfo{author}{Cliffe, K.~A.} \& \bibinfo{author}{Mullin, T.}
\newblock \bibinfo{title}{Bifurcation phenomena in taylor-couette flow in a
  very short annulus}.
\newblock \emph{\bibinfo{journal}{J. Fluid Mech.}}
  \textbf{\bibinfo{volume}{191}}, \bibinfo{pages}{1--18}
  (\bibinfo{year}{1988}).

\bibitem{PSL:1991}
\bibinfo{author}{Pfister, G.}, \bibinfo{author}{Schulz, A.} \&
  \bibinfo{author}{B., L.}
\newblock \bibinfo{title}{Bifurcations and a route to chaos of an
  one-vortex-state in taylor-couette flow}.
\newblock \emph{\bibinfo{journal}{Eur. J. Mech. B}}
  \textbf{\bibinfo{volume}{10}}, \bibinfo{pages}{247--252}
  (\bibinfo{year}{1991}).

\bibitem{PBE:1992}
\bibinfo{author}{Pfister, G.}, \bibinfo{author}{Buzug, T.} \&
  \bibinfo{author}{Enge, N.}
\newblock \bibinfo{title}{Characterization of experimental time series from
  taylor-couette flow}.
\newblock \emph{\bibinfo{journal}{Physica D}} \textbf{\bibinfo{volume}{58}},
  \bibinfo{pages}{441--454} (\bibinfo{year}{1992}).

\bibitem{Cli:1983}
\bibinfo{author}{Cliffe, K.~A.}
\newblock \bibinfo{title}{Numerical calculations of two-cell and single-cell
  taylor flows}.
\newblock \emph{\bibinfo{journal}{J. Fluid Mech.}}
  \textbf{\bibinfo{volume}{135}}, \bibinfo{pages}{219--233}
  (\bibinfo{year}{1983}).

\bibitem{SPT:2003}
\bibinfo{author}{Schulz, A.}, \bibinfo{author}{Pfister, G.} \&
  \bibinfo{author}{Tavener, S.~J.}
\newblock \bibinfo{title}{The effect of outer cylinder rotation on
  taylor–couette flow at small aspect}.
\newblock \emph{\bibinfo{journal}{Phys. Fluids}} \textbf{\bibinfo{volume}{15}},
  \bibinfo{pages}{417--425} (\bibinfo{year}{1991}).

\bibitem{NaTo:1996}
\bibinfo{author}{Nakamura, I.} \& \bibinfo{author}{Toya, Y.}
\newblock \bibinfo{title}{Existence of extra vortex and twin vortex of
  anomalous mode in taylor vortex flow with a small aspect ratio}.
\newblock \emph{\bibinfo{journal}{Acta Mech.}} \textbf{\bibinfo{volume}{117)}},
  \bibinfo{pages}{33--46} (\bibinfo{year}{1996}).

\bibitem{BSP:1992}
\bibinfo{author}{Buzug, T.}, \bibinfo{author}{von Stamm, J.} \&
  \bibinfo{author}{Pfister, G.}
\newblock \bibinfo{title}{Characterization of experimental time series from
  taylor-couette flow}.
\newblock \emph{\bibinfo{journal}{Physica A}} \textbf{\bibinfo{volume}{191}},
  \bibinfo{pages}{559} (\bibinfo{year}{1992}).

\bibitem{FWTN:2002}
\bibinfo{author}{Furukawa, H.}, \bibinfo{author}{Watanabe, T.},
  \bibinfo{author}{Toya, Y.} \& \bibinfo{author}{I., N.}
\newblock \bibinfo{title}{Flow pattern exchange in the taylor-couette system
  with a very small aspect ratio}.
\newblock \emph{\bibinfo{journal}{Phys.~Rev.~E}} \textbf{\bibinfo{volume}{65}},
  \bibinfo{pages}{036306} (\bibinfo{year}{1992}).

\bibitem{ALD:2012}
\bibinfo{author}{Altmeyer, S.}, \bibinfo{author}{Lopez, J.} \&
  \bibinfo{author}{Do, Y.}
\newblock \bibinfo{title}{Influence of an inhomogeneous internal magnetic field
  on the flow dynamics of ferrofluid between differentially rotating
  cylinders}.
\newblock \emph{\bibinfo{journal}{Phys.~Rev.~E}} \textbf{\bibinfo{volume}{85}},
  \bibinfo{pages}{066314} (\bibinfo{year}{2012}).

\bibitem{GSS88}
\bibinfo{author}{Golubitsky, M.}, \bibinfo{author}{Stewart, I.} \&
  \bibinfo{author}{Schaeffer, D.}
\newblock \emph{\bibinfo{title}{Singularities and Groups in Bifurcation Theory
  II}} (\bibinfo{publisher}{Springer}, \bibinfo{address}{New York},
  \bibinfo{year}{1988}).

\bibitem{GL88}
\bibinfo{author}{Golubitsky, M.} \& \bibinfo{author}{Langford, W.~F.}
\newblock \bibinfo{title}{Pattern formation and bistability in flow between
  counterrotating cylinders}.
\newblock \emph{\bibinfo{journal}{Physica D}} \textbf{\bibinfo{volume}{32}},
  \bibinfo{pages}{362--392} (\bibinfo{year}{1988}).

\bibitem{WeLu:1998}
\bibinfo{author}{Wereley, S.~T.} \& \bibinfo{author}{Lueptow, R.~M.}
\newblock \bibinfo{title}{Spatio-temporal character of non-wavy and wavy
  taylor-couette flow}.
\newblock \emph{\bibinfo{journal}{J.~Fluid~Mech.}}
  \textbf{\bibinfo{volume}{364}}, \bibinfo{pages}{59 -- 80}
  (\bibinfo{year}{1998}).

\bibitem{CHSAML2009}
\bibinfo{author}{Hoffmann, C.}, \bibinfo{author}{Altmeyer, S.},
  \bibinfo{author}{Pinter, A.} \& \bibinfo{author}{L\"ucke, M.}
\newblock \bibinfo{title}{Transitions between taylor vortices and spirals via
  wavy taylor vortices and wavy spirals}.
\newblock \emph{\bibinfo{journal}{New~J.~Phys.}} \textbf{\bibinfo{volume}{11}},
  \bibinfo{pages}{053002} (\bibinfo{year}{2009}).

\bibitem{MSL:2014}
\bibinfo{author}{Martinand, D.}, \bibinfo{author}{Serre, E.} \&
  \bibinfo{author}{Lueptow, R.}
\newblock \bibinfo{title}{Mechanisms for the transition to waviness for taylor
  vortices}.
\newblock \emph{\bibinfo{journal}{Phys.~Fluids}} \textbf{\bibinfo{volume}{26}},
  \bibinfo{pages}{094102} (\bibinfo{year}{2014}).

\bibitem{ADML2012}
\bibinfo{author}{Altmeyer, S.}, \bibinfo{author}{Do, Y.},
  \bibinfo{author}{Marquez, F.} \& \bibinfo{author}{Lopez, J.~M.}
\newblock \bibinfo{title}{Symmetry-breaking hopf bifurcations to 1-, 2-, and
  3-tori in small-aspect-ratio counterrotating taylor-couette flow}.
\newblock \emph{\bibinfo{journal}{Phys.~Rev.~E}} \textbf{\bibinfo{volume}{86}}
  (\bibinfo{year}{2012}).

\bibitem{SACHALML2010}
\bibinfo{author}{Altmeyer, S.}, \bibinfo{author}{Hoffmann, C.},
  \bibinfo{author}{M., A.~L.} \& \bibinfo{author}{L\"ucke}.
\newblock \bibinfo{title}{Influence of homogeneous magnetic fields on the flow
  of a ferrofluid in the taylor-couette system}.
\newblock \emph{\bibinfo{journal}{Phys.~Rev.~E}} \textbf{\bibinfo{volume}{82}},
  \bibinfo{pages}{016321} (\bibinfo{year}{2010}).

\bibitem{DiSw:1985}
\bibinfo{author}{DiPrima, R.~C.} \& \bibinfo{author}{Swinney, H.~L.}
\newblock \bibinfo{title}{Instabilities and transition in flow between
  concentric rotating cylinders}.
\newblock In \bibinfo{editor}{Swinney, H.~L.} \& \bibinfo{editor}{Gollub,
  J.~G.} (eds.) \emph{\bibinfo{booktitle}{Hydrodynamic Instabilities and the
  Transition to Turbulence}}, no.~\bibinfo{number}{45} in
  \bibinfo{series}{Topics in Applied Physics} (\bibinfo{publisher}{Springer},
  \bibinfo{address}{Berlin}, \bibinfo{year}{1985}).

\bibitem{ALS:1986}
\bibinfo{author}{Andereck, C.~D.}, \bibinfo{author}{Liu, S.~S.} \&
  \bibinfo{author}{Swinney, H.~L.}
\newblock \bibinfo{title}{Flow regimes in a circular couette system with
  independently rotating cylinders}.
\newblock \emph{\bibinfo{journal}{J. Fluid Mech.}}
  \textbf{\bibinfo{volume}{164}}, \bibinfo{pages}{155--183}
  (\bibinfo{year}{1986}).

\bibitem{Nag:1988}
\bibinfo{author}{Nagata, M.}
\newblock \bibinfo{title}{On wavy instabilities of the taylor-vortex flow
  between corotating cylinders}.
\newblock \emph{\bibinfo{journal}{J. Fluid Mech.}}
  \textbf{\bibinfo{volume}{88}}, \bibinfo{pages}{585--598}
  (\bibinfo{year}{1988}).

\bibitem{MuLi:2001}
\bibinfo{author}{M\"{u}ller, H.~W.} \& \bibinfo{author}{Liu, M.}
\newblock \bibinfo{title}{Structure of ferrofluid dynamics}.
\newblock \emph{\bibinfo{journal}{Phys.~Rev.~E}} \textbf{\bibinfo{volume}{64}},
  \bibinfo{pages}{061405} (\bibinfo{year}{2001}).

\bibitem{Lan:1905}
\bibinfo{author}{Langevin, P.}
\newblock \bibinfo{title}{Magn\'etisme et th\'eorie des \'electrons}.
\newblock \emph{\bibinfo{journal}{Annales de Chemie et de Physique}}
  \textbf{\bibinfo{volume}{5}}, \bibinfo{pages}{70--127}
  (\bibinfo{year}{1905}).

\bibitem{EMWKL:2000}
\bibinfo{author}{Embs, J.}, \bibinfo{author}{M\"uller, H.~W.},
  \bibinfo{author}{Wagner, C.}, \bibinfo{author}{Knorr, K.} \&
  \bibinfo{author}{L\"ucke, M.}
\newblock \bibinfo{title}{Measuring the rotational viscosity of ferrofluids
  without shear flow}.
\newblock \emph{\bibinfo{journal}{Phys.~Rev.~E}} \textbf{\bibinfo{volume}{61}},
  \bibinfo{pages}{R2196--R2199} (\bibinfo{year}{2000}).

\bibitem{Nik:1987}
\bibinfo{author}{Niklas, M.}
\newblock \bibinfo{title}{Influence of magnetic fields on {T}aylor vortex
  formation in magnetic fluids}.
\newblock \emph{\bibinfo{journal}{Z. Phys. B}} \textbf{\bibinfo{volume}{68}},
  \bibinfo{pages}{493} (\bibinfo{year}{1987}).

\bibitem{NML:1989}
\bibinfo{author}{Niklas, M.}, \bibinfo{author}{M\"uller-Krumbhaar, H.} \&
  \bibinfo{author}{L\"ucke, M.}
\newblock \bibinfo{title}{Taylor--vortex flow of ferrofluids in the presence of
  general magnetic fields}.
\newblock \emph{\bibinfo{journal}{J.~Magn.~Magn.~Mater.}}
  \textbf{\bibinfo{volume}{81}}, \bibinfo{pages}{29} (\bibinfo{year}{1989}).

\bibitem{OdMu:2002}
\bibinfo{author}{Odenbach, S.} \& \bibinfo{author}{M\"uller, H.~W.}
\newblock \bibinfo{title}{Stationary off-equilibrium magnetization in
  ferrofluids under rotational and elongational flow}.
\newblock \emph{\bibinfo{journal}{Phys.~Rev.~Lett.}}
  \textbf{\bibinfo{volume}{89}}, \bibinfo{pages}{037202}
  (\bibinfo{year}{2002}).

\end{thebibliography}

\section*{Acknowledgement}
Y.D. was supported by Basic Science Research Program 
of the Ministry of Education, Science and Technology.
Y.C.L. was supported by AFOSR under Grant No.~FA9550-15-1-0151.
Y.C.L. would also like to acknowledge support from the Vannevar Bush Faculty Fellowship program sponsored by the Basic Research Office of the Assistant Secretary of Defense for Research and Engineering and funded by the Office of Naval Research through Grant No.~N00014-16-1-2828.

\section*{Author contributions}
S.A. devised the research project and performed numerical simulations 
and analyzed the results. All wrote the paper.

\section*{Additional information}

{\bf Competing financial interests}:
The authors declare no competing financial interests.

\clearpage

\clearpage

\end{document}